\documentclass[superscriptaddress,twocolumn,english,floatfix,aps,pra,amsmath,amssymb,longbibliography]{revtex4-2}
\usepackage[T1]{fontenc}
\usepackage[utf8]{inputenc}
\usepackage{babel}
\usepackage{comment}
\usepackage{color}
\usepackage{times}
\usepackage{epsfig}
\usepackage{subfigure}
\usepackage{amsmath}
\usepackage{amssymb}
\usepackage{graphicx}
\usepackage[colorlinks=true]{hyperref}
\hypersetup{citecolor=blue,linkcolor=blue,urlcolor=blue}

\begin{document}

\title{Regimes of the lateral van der Waals force in the presence of dielectrics}

\author{Lucas Queiroz}
\email{lucas.silva@icen.ufpa.br}
\affiliation{Faculdade de F\'{i}sica, Universidade Federal do Par\'{a}, 66075-110, Bel\'{e}m, Par\'{a}, Brazil}

\author{Edson C. M. Nogueira}
\email{edson.moraes.nogueira@icen.ufpa.br}
\affiliation{Faculdade de F\'{i}sica, Universidade Federal do Par\'{a}, 66075-110, Bel\'{e}m, Par\'{a}, Brazil}

\author{Danilo T. Alves}
\email{danilo@ufpa.br}
\affiliation{Faculdade de F\'{i}sica, Universidade Federal do Par\'{a}, 66075-110, Bel\'{e}m, Par\'{a}, Brazil}
\affiliation{Centro de F\'{i}sica, Universidade do Minho, P-4710-057, Braga, Portugal}

\date{\today}

\begin{abstract}
In a recent paper, it was shown that, under the action of the lateral van der Waals (vdW) force due to a perfectly conducting corrugated surface, a neutral anisotropic polarizable particle in vacuum can be attracted not only to the nearest corrugation peak, but also to a valley, or an intermediate point between a peak and a valley, with such behaviors called peak, valley and intermediate regimes, respectively.
In the present paper, we investigate how these regimes are affected by the consideration of two non-dispersive semi-infinite dielectrics $\epsilon_{1}$ and $\epsilon_{2}$, separated by a corrugated interface.
Specifically, we study the vdW interaction between a neutral anisotropic polarizable particle, embedded in the dielectric $\epsilon_{2}$, and the dielectric $\epsilon_{1}$.
We show that when $\epsilon_{2}<\epsilon_{1}$
the peak, valley and intermediate regimes have, unless numerical factors, behaviors similar to those found
for the situation where the particle is in vacuum and interacting with a conducting medium.
For the case $\epsilon_{2}>\epsilon_{1}$, one might expect a mere permute between the peak and valley regimes,
in comparison to the case $\epsilon_{2}<\epsilon_{1}$.
Surprisingly, we find that when $\epsilon_{2}>\epsilon_{1}$ the regimes exhibit a very different and nontrivial behavior.
Moreover, we show that similar regimes arise in the classical case involving a neutral polarized particle.
The description of how the peak, valley and intermediate regimes are affected by the presence of dielectrics may be relevant for a better understanding of the interaction between anisotropic particles and corrugated surfaces.
\end{abstract}

\maketitle

\section{Introduction}
\label{sec-intro}

Casimir-Polder/van der Waals (CP/vdW) dispersion forces \cite{Casimir-Polder-PhysRev-1948} have important consequences in many areas of science \cite{Israelachvili-PRSL-1972, Milonni-QuantumVacuum-1994, Bordag-Book-2009, Israelachvili-Book-2011, Buhmann-DispersionForces-I, Buhmann-DispersionForces-II, Passante-Symmetry-2018,Dimopoulos-PRD-2003,Parsegian-Book-2006,Woods-RMP-2016}, and the growing progress in their experimental measurement has opened possibilities for many applications in micro and nanotechnology \cite{Ball-Nature-2007, Rodriguez-Capasso-PRL-2010, Rodriguez-NaturePhotonics-2011, Keil-JourModOpt-2016,Laliotis-Arxiv-2021}.
In the context of the CP/vdW interaction, the consideration of anisotropic particles has increased, 
since they are involved with nontrivial behaviors.
Among them, one has the prediction of a repulsive CP/vdW force involving these particles \cite{Levin-PRL-2010, Eberlein-PRA-2011, Buhmann-IJMPA-2016, Abrantes-PRA-2018, Venkataram-PRA-2020, Marchetta-PRA-2021} and the phenomenon known as the Casimir torque \cite{Bimonte-PRD-2015, Thiyam-PRA-2015, Gangaraj-PRB-2018, Antezza-PRB-2020}.
Furthermore, making analytical calculations that go beyond the proximity force approximation (PFA),
in Ref. \cite{Nogueira-PRA-2021} the present authors predicted that, under the action of the lateral vdW force, an anisotropic particle can be attracted not only toward the corrugation peaks (as found until that moment in the literature \cite{Bezerra-PRA-2000, Dalvit-PRL-2008, Buhmann-IJMPA-2016}), but also to the nearest valley, or to an intermediate point between a peak and valley, with such behaviors called 
as peak, valley and intermediate regimes, respectively.

In the present paper, taking as basis the analytical perturbative approach discussed 
in Refs. \cite{Clinton-PRB-1985,Alves-PRB-2019,Nogueira-PRA-2021},
we investigate how the peak, valley and intermediate regimes are affected by the consideration of dielectric media.
We consider a neutral anisotropic polarizable particle in a semi-infinite medium with dielectric constant $\epsilon_{2}$, separated by a corrugated surface of another semi-infinite medium with dielectric constant $\epsilon_{1}$.
We consider, as done in Refs. \cite{Reyes-PRA-2009, Eberlein-PRA-2010},
non-dispersive dielectric media in our calculations.
As discussed along the present work, this consideration enable us to write
simpler formulas, which already provide a first estimate about the non-trivial behavior of the peak, valley
and intermediate regimes in the presence of dielectric media.

The paper is organized as follows. 
In Sec. \ref{sec-used-approach}, we present the approach used in this work to write formulas for the classical and the quantum (vdW) interaction between a neutral particle, embedded in a dielectric $\epsilon_{2}$ and in the presence of another dielectric $\epsilon_{1}$, separated of $\epsilon_{2}$ by a corrugated interface.
In Sec. \ref{sec-classical-case}, we investigate the classical interaction between a neutral polarized particle, embedded in the dielectric $\epsilon_{2}$, and the dielectric $\epsilon_{1}$, and we also apply our formulas to the case of a sinusoidal corrugation.
In Sec. \ref{sec-vdw}, we compute the vdW interaction involving a neutral polarizable particle, in the same configuration of dielectrics of the previous section, and we also apply our formulas to the case of a sinusoidal corrugation.
In Sec. \ref{sec-final}, we present our final comments as well as discuss some implications of our results.

\section{Our Approach}
\label{sec-used-approach}

The first part of our approach is to obtain, using the analytical approach found in Refs. \cite{Clinton-PRB-1985,Alves-PRB-2019}, the Green function related to Poisson's equation of a point charge put in a semi-infinite dielectric $\epsilon_{2}$, separated by a corrugated surface of another semi-infinite dielectric $\epsilon_{1}$. 
The second part is to use this Green function in the formula obtained in Ref. \cite{Eberlein-PRA-2007} to compute the vdW interaction between the polarizable particle and the medium $\epsilon_{1}$.

\begin{figure}[h]
\centering
\subfigure[]{\label{fig:carga-2-meios-plano}\epsfig{file=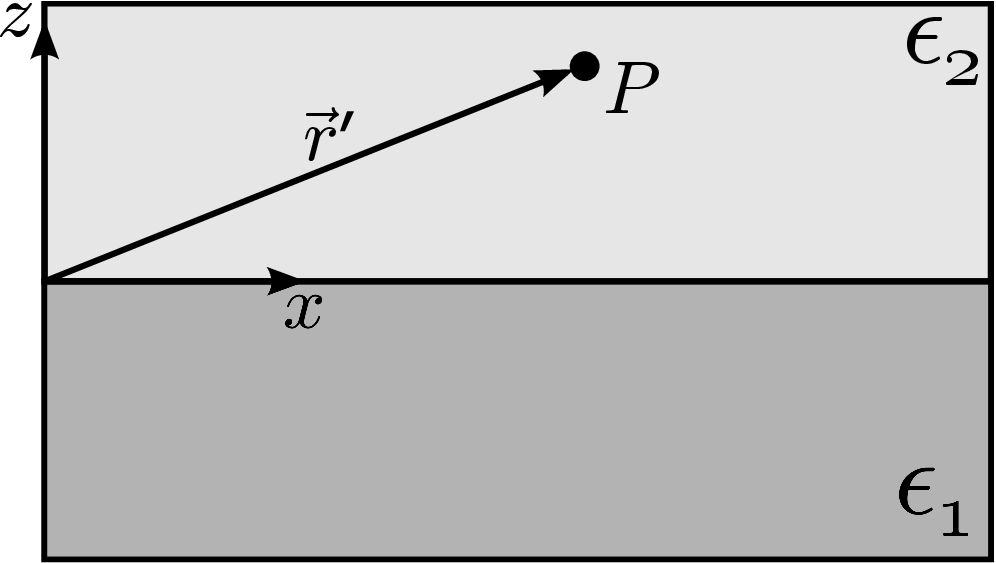, width=0.45 \linewidth}}
\hspace{4mm}
\subfigure[]{\label{fig:carga-2-meios-corrug}\epsfig{file=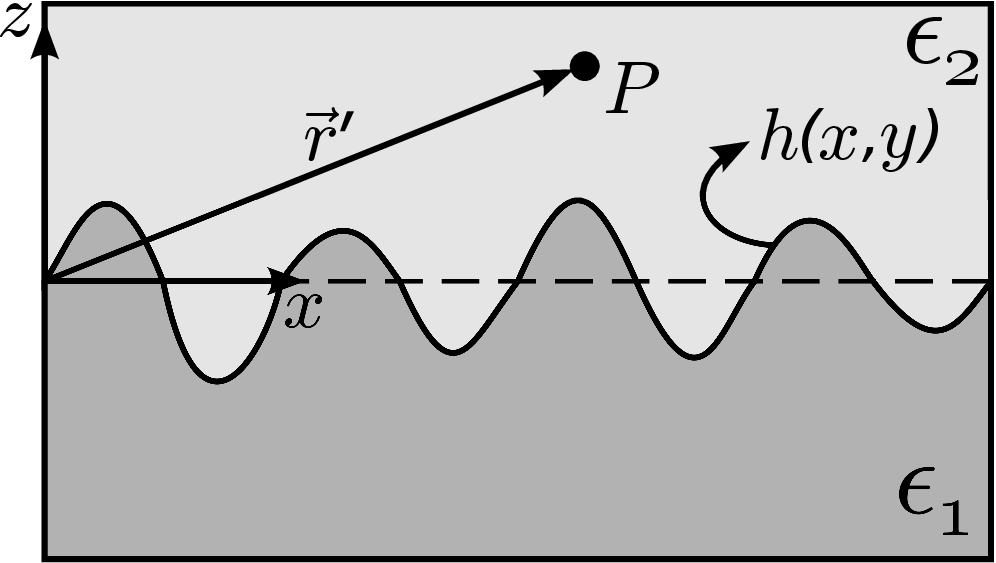, width=0.45 \linewidth}}
\caption{
Illustration of two non-dispersive semi-infinite dielectrics $\epsilon_{1}$ and $\epsilon_{2}$, with a particle (represented by the point $P$) embedded in $\epsilon_{2}$ and located at ${\bf r}^{\prime}=x^\prime\hat{{\bf x}}+y^\prime\hat{{\bf y}}+z^\prime\hat{{\bf z}}$ (with $z^\prime>0$). 
The point $P$ represents a charge, a dipole moment or a polarizable particle, in Secs. \ref{sec-used-approach}, \ref{sec-classical-case} and \ref{sec-vdw}, respectively.
In (a) the dielectrics are separated by a planar interface, whereas in (b) they are separated by a corrugated one, whose corrugation profile is described by $z=h(x,y)$.
}
\label{fig:carga-2-meios}
\end{figure}

In the literature (for instance, in Ref. \cite{Jackson-Electrodynamics-1998}) is found the Green function for the problem of two semi-infinite dielectrics, $\epsilon_1$ and $\epsilon_2$,
separated by a planar surface at $z=0$, with a point charge $Q$ embedded in the dielectric $\epsilon_2$
[see Fig. \ref{fig:carga-2-meios-plano}].
Here, we consider these two non-dispersive dielectrics separated by a general corrugated interface, as illustrated in Fig. \ref{fig:carga-2-meios-corrug}.
Mathematically, the dielectric constant of this system can be described by
\begin{equation}
\epsilon\left({\bf r}\right)=\begin{cases}
\epsilon_{1}, & z< h\left({\bf r}_{||}\right)\\
\epsilon_{2}, & z> h\left({\bf r}_{||}\right)
\end{cases}, \label{eq:epsilon}
\end{equation}
where $h\left({\bf r}_{\parallel}\right)$ describes a suitable modification of a planar interface, between the dielectrics $\epsilon_{1}$ and $\epsilon_{2}$, at $z=0$ [see Fig. \ref{fig:carga-2-meios-corrug}].
For the purposes of this paper, it is interesting to write Eq. \eqref{eq:epsilon} in terms of Heaviside functions as
\begin{equation}
\epsilon\left(\bf{r}\right)=\epsilon_{1}\theta\left[-z+ \eta h\left({\bf r}_{\parallel}\right)\right]+\epsilon_{2}\theta\left[z- \eta h\left({\bf r}_{\parallel} \right)\right], \label{eq:epsilon-heaviside}
\end{equation}
where we introduced an arbitrary dimensionless auxiliary parameter $\eta$, with $0 \leq \eta \leq 1$, such that, for $\eta=0$ we recover a planar interface between $\epsilon_{1}$ and $\epsilon_{2}$, and for $\eta = 1$ we have the actual corrugated interface described by $h\left({\bf r}_{\parallel}\right)$.
Note that for $h({\bf r}_{\parallel})=0$ Eq. \eqref{eq:epsilon-heaviside} describes the case where $\epsilon_1$ and $\epsilon_2$ are separated by a planar interface at $z=0$.
Performing an expansion in powers of $\eta h$, we can write
\begin{eqnarray}
\nonumber \epsilon({\bf r}) & = & \epsilon_{1}\theta(-z)+\epsilon_{2}\theta(z)+(\epsilon_{1}-\epsilon_{2})\delta(z)\eta h({\bf r}_{\parallel}) \\
& & -\frac{1}{2}(\epsilon_{1}-\epsilon_{2})\delta^{\prime}(z)\eta^2 h^{2}({\bf r}_{\parallel})+... .
\label{eq:epsilon-heaviside-2}
\end{eqnarray}

For a charge $Q$ put in the dielectric $\epsilon_{2}$, and located at the position ${\bf r}^{\prime}={\bf r}_{||}^{\prime}+z^\prime\hat{{\bf z}}$
(with $z^\prime>0$ and ${\bf r}_{||}^\prime=x^\prime\hat{{\bf x}}+y^\prime\hat{{\bf y}}$), one has the
Poisson equation
\begin{equation}
\boldsymbol{\nabla}\cdot\left[\epsilon\left({\bf r}\right)\boldsymbol{\nabla}{\bf \phi\left({\bf r}\right)}\right]=-\frac{4\pi Q}{\epsilon_0}\delta\left({\bf r}-{\bf r}^{\prime}\right), \label{eq:poisson-coordenadas}
\end{equation}
where the potential $\phi$ can be written as 
$\phi\left({\bf r}, {\bf r}^{\prime}\right)=\frac{Q}{\epsilon_0}G\left({\bf r},{\bf r}^{\prime}\right)$, with
$G\left({\bf r},{\bf r}^{\prime}\right)$ being the Green function of the problem.
Following Clinton, Esrik, and Sacks \cite{Clinton-PRB-1985}, we look for a perturbative approximate solution of $G$ with $\text{max}|h({\bf r}_{||})|=a\ll z^\prime$, which, up to first order, can be written as
\begin{equation}
G\left(\textbf{r},\textbf{r}^{\prime}\right) \approx G^{(0)}\left(\textbf{r},\textbf{r}^{\prime}\right) + \eta
G^{(1)}\left(\textbf{r},\textbf{r}^{\prime}\right), \label{eq:g-coordenadas}
\end{equation}
where $G^{(0)}\left(\textbf{r},\textbf{r}^{\prime}\right)$ is the unperturbed solution related to 
the situation shown in Fig. \ref{fig:carga-2-meios-plano}, and $G^{(1)}\left(\textbf{r},\textbf{r}^{\prime}\right)$ 
is the first-order correction to $G^{(0)}$ due to the surface corrugation.
In order to find $G$ by solving Eq. \eqref{eq:poisson-coordenadas}, it is convenient to work in the reduced Fourier space, so that we have to perform, in Eq. \eqref{eq:poisson-coordenadas}, a Fourier transformation in the $x$ and $y$ coordinates. 
Thus, let us define the Fourier transformation of a function $f({\bf r}_{||})$ as
\begin{equation}
f\left({\bf q}\right)=\int d^{2}{\bf r}_{\parallel}e^{-i{\bf q}\cdot{\bf r}_{\parallel}}f\left({\bf r}_{\parallel}\right),
\end{equation}
and its inverse transformation as
\begin{equation}
f\left({\bf r}_{\parallel}\right)=\int\frac{d^{2}{\bf q}}{\left(2\pi\right)^{2}}e^{i{\bf q}\cdot{\bf r}_{\parallel}}f\left({\bf q}\right),
\end{equation}
where ${\bf q} = q_x\hat{{\bf x}}+q_y\hat{{\bf y}}$, and, applying this transformation in Eq. \eqref{eq:poisson-coordenadas}, we obtain
\begin{align}
\nonumber \int\frac{d^{2}{\bf q}^{\prime}}{\left(2\pi\right)^{2}}&\{\frac{\partial}{\partial z}[\epsilon({\bf q}-{\bf q}^{\prime},z)\frac{\partial}{\partial z}{\cal G}({\bf q}^{\prime},z,{\bf r^{\prime}})]-({\bf q}\cdot{\bf q}^{\prime})\\
\times\epsilon({\bf q}&-{\bf q}^{\prime},z){\cal G}({\bf q}^{\prime},z,{\bf r^{\prime}})\}=-4\pi e^{-i{\bf q}\cdot{\bf r}_{\parallel}^{\prime}}\delta(z-z^{\prime}), \label{eq:poisson-fourier}
\end{align}
where ${\cal G}({\bf q}^{\prime},z,{\bf r^{\prime}})$ is the Fourier transformation of $G\left({\bf r},{\bf r^{\prime}}\right)$. From Eq. \eqref{eq:g-coordenadas}, we can write
\begin{equation}
{\cal G}({\bf q}^{\prime},z,{\bf r^{\prime}})={\cal G}^{(0)}({\bf q}^{\prime},z,{\bf r^{\prime}})+\eta{\cal G}^{(1)}({\bf q}^{\prime},z,{\bf r^{\prime}}), \label{eq:g-fourier}
\end{equation}
and 
\begin{align}
\epsilon({\bf q}-{\bf q}^{\prime},z)= & (2\pi)^{2}[\epsilon_{1}\theta(-z)+\epsilon_{2}\theta(z)]\delta({\bf q}-{\bf q}^{\prime})\nonumber \\
& +(\epsilon_{1}-\epsilon_{2})\delta(z)\eta h_{(1)}({\bf q}-{\bf q}^{\prime})\nonumber \\
& -\frac{1}{2}(\epsilon_{1}-\epsilon_{2})\delta^{\prime}(z)\eta^2 h_{(2)}({\bf q}-{\bf q}^{\prime})+... \label{eq:epsilon-fourier}
\end{align}
with 
\begin{equation}
h_{(m)}({\bf q}-{\bf q}^{\prime})=\int d^{2}{\bf r}_{\parallel}[h({\bf r}_{\parallel})]^{m}e^{-i({\bf q}-{\bf q}^{\prime})\cdot{\bf r}_{\parallel}},
\end{equation}
where $m \in \mathbb{N}^* $. 
Substituting Eqs. \eqref{eq:g-fourier} and \eqref{eq:epsilon-fourier} in Eq. \eqref{eq:poisson-fourier}, and since $\eta$ is an arbitrary parameter, we obtain differential equations specifically to ${\cal G}^{(0)}$ and ${\cal G}^{(1)}$, which are given by
\begin{align}
\frac{\partial}{\partial z}[(\epsilon_{1}\theta(-z)+\epsilon_{2}\theta(z))&\frac{\partial}{\partial z}{\cal G}^{(0)}({\bf q},z,{\bf r^{\prime}})] \nonumber \\
-|{\bf q}|^{2}(\epsilon_{1}\theta(-z)+\epsilon_{2}&\theta(z)){\cal G}^{(0)}({\bf q},z,{\bf r^{\prime}})  \nonumber \\
& =-4\pi e^{-i{\bf q}\cdot{\bf r}_{\parallel}^{\prime}}\delta(z-z^{\prime}), \label{eq:g0-fourier}
\end{align}
and 
\begin{align}
(\epsilon_{1}-\epsilon_{2}) & \int\frac{d^{2}{\bf q}^{\prime}}{\left(2\pi\right)^{2}}\{\frac{\partial}{\partial z}[\delta(z)h_{(1)}({\bf q}-{\bf q}^{\prime})\frac{\partial}{\partial z}{\cal G}^{(0)}({\bf q}^{\prime},z,{\bf r^{\prime}})]\nonumber \\
&-({\bf q}\cdot{\bf q}^{\prime})\delta(z)h_{(1)}({\bf q}-{\bf q}^{\prime}){\cal G}^{(0)}({\bf q}^{\prime},z,{\bf r^{\prime}})\}\nonumber \\
=&-\frac{\partial}{\partial z}[(\epsilon_{1}\theta(-z)+\epsilon_{2}\theta(z))\frac{\partial}{\partial z}{\cal G}^{(1)}({\bf q},z,{\bf r^{\prime}})]\nonumber \\
&+|{\bf q}|^{2}(\epsilon_{1}\theta(-z)+\epsilon_{2}\theta(z)){\cal G}^{(1)}({\bf q},z,{\bf r^{\prime}}), \label{eq:g1-fourier}
\end{align}
respectively.

It is required that ${\cal G}({\bf q}^{\prime},z,{\bf r^{\prime}})$ be continuous for all values of $z$, including in the interface between the dielectrics $\epsilon_{1}$ and $\epsilon_{2}$, in which we have
\begin{equation}
\lim_{\zeta\to0^{+}}{\cal G}({\bf q}^{\prime},h({\bf r}_{\parallel})+\zeta,{\bf r^{\prime}})=\lim_{\zeta\to0^{+}}{\cal G}({\bf q}^{\prime},h({\bf r}_{\parallel})-\zeta,{\bf r^{\prime}}).
\label{eq:condicao-continuidade-g}
\end{equation}
Substituting in this equation the approximate solution for ${\cal G}$, given in Eq. \eqref{eq:g-fourier}, and expanding it in powers of $\eta h$, we obtain conditions for ${\cal G}^{(0)}$ and ${\cal G}^{(1)}$ in the interface.  These are given by
\begin{equation}
{\cal G}^{(0)}({\bf q},0^{-},{\bf r^{\prime}})={\cal G}^{(0)}({\bf q},0^{+},{\bf r^{\prime}}), \label{eq:g0-condicao}
\end{equation}
and
\begin{align}
{\cal G}^{(1)}({\bf q},0^{+},{\bf r^{\prime}})-{\cal G}^{(1)}({\bf q},0^{-},{\bf r^{\prime}}) & =-\int\frac{d^{2}{\bf q}^{\prime}}{\left(2\pi\right)^{2}}h_{(1)}\left({\bf q}-{\bf q}^{\prime}\right)\nonumber \\
\times\{[\frac{\partial}{\partial z}{\cal G}^{(0)}({\bf q}^{\prime},z,{\bf r^{\prime}})]_{z=0^{+}} & -[\frac{\partial}{\partial z}{\cal G}^{(0)}({\bf q}^{\prime},z,{\bf r^{\prime}})]_{z=0^{-}}\}, \label{eq:g1-condicao}
\end{align}
respectively.
To obtain the solutions for ${\cal G}^{(0)}$ and ${\cal G}^{(1)}$, it is also necessary to investigate the discontinuity condition of their $z$-derivative calculated at $z=0$.
This can be done by integrating Eqs. \eqref{eq:g0-fourier} and \eqref{eq:g1-fourier} around $z=0$, so that, we obtain
\begin{equation}
\epsilon_{1}[\frac{\partial}{\partial z}{\cal G}^{(0)}({\bf q},z,{\bf r^{\prime}})]_{z=0^{-}}=\epsilon_{2}[\frac{\partial}{\partial z}{\cal G}^{(0)}({\bf q},z,{\bf r^{\prime}})]_{z=0^{+}}, \label{eq:g0-condicao-derivada}
\end{equation}
for the $z$-derivative of ${\cal G}^{(0)}$, and
\begin{align}
\epsilon_{2}[&\frac{\partial}{\partial z}{\cal G}^{(1)}({\bf q},z,{\bf r^{\prime}})]_{z=0^{+}}-\epsilon_{1}[\frac{\partial}{\partial z}{\cal G}^{(1)}({\bf q},z,{\bf r^{\prime}})]_{z=0^{-}}\nonumber \\
=&(\epsilon_{1}-\epsilon_{2})\int\frac{d^{2}{\bf q}^{\prime}}{\left(2\pi\right)^{2}}\left({\bf q}\cdot{\bf q}^{\prime}\right)h_{(1)}\left({\bf q}-{\bf q}^{\prime}\right){\cal G}^{(0)}({\bf q}^{\prime},0,{\bf r^{\prime}}), \label{eq:g1-condicao-derivada}
\end{align}
for the $z$-derivative of ${\cal G}^{(1)}$.

The solution for Eq. \eqref{eq:g0-fourier}, with boundary conditions given in Eqs. \eqref{eq:g0-condicao} and \eqref{eq:g0-condicao-derivada}, and the condition that ${\cal G}^{(0)}$ goes to zero for large distances, can be written as
\begin{equation}
{\cal G}^{(0)}({\bf q},z,{\bf r^{\prime}})=\begin{cases}
{\cal G}_{<}^{(0)}({\bf q},z,{\bf r^{\prime}}), & z<0\\
{\cal G}_{>}^{(0)}({\bf q},z,{\bf r^{\prime}}), & z>0
\end{cases}, \label{eq:sol-g0}
\end{equation}
where the subscripts ``$<$'' and ``$>$'' are related to the solutions for the regions $z<0$ and $z>0$, respectively, which are given by
\begin{equation}
{\cal G}_{<}^{(0)}({\bf q},z,{\bf r^{\prime}})=\frac{4\pi e^{-i{\bf q}\cdot{\bf r}_{\parallel}^{\prime}}}{(\epsilon_{1}+\epsilon_{2})|{\bf q}|}e^{-|{\bf q}|(z^{\prime}-z)}, \label{eq:sol-g0-menor}
\end{equation}
and 
\begin{equation}
{\cal G}_{>}^{(0)}({\bf q},z,{\bf r^{\prime}})=\frac{2\pi e^{-i{\bf q}\cdot{\bf r}_{\parallel}^{\prime}}}{\epsilon_{2}|{\bf q}|}[e^{-|{\bf q}||z-z^{\prime}|}-\frac{\epsilon_{1}-\epsilon_{2}}{\epsilon_{1}+\epsilon_{2}}e^{-|{\bf q}|(z+z^{\prime})}]. \label{eq:sol-g0-maior}
\end{equation}
In the same way, the solution for Eq. \eqref{eq:g1-fourier}, with boundary conditions given in Eqs. \eqref{eq:g1-condicao} and \eqref{eq:g1-condicao-derivada}, and the condition that ${\cal G}^{(1)}$ goes to zero for large distances, can be written as
\begin{equation}
{\cal G}^{(1)}({\bf q},z,{\bf r^{\prime}})=\begin{cases}
{\cal G}_{<}^{(1)}({\bf q},z,{\bf r^{\prime}}), & z<0\\
{\cal G}_{>}^{(1)}({\bf q},z,{\bf r^{\prime}}), & z>0
\end{cases}, \label{eq:sol-g1}
\end{equation}
where
\begin{equation}
{\cal G}_{<}^{(1)}({\bf q},z,{\bf r^{\prime}})=e^{|{\bf q}|z}\int\frac{d^{2}{\bf q}^{\prime}}{\left(2\pi\right)^{2}}p({\bf q},{\bf q}^{\prime},{\bf r^{\prime}})\left[1-\frac{{\bf q}^{\prime}\cdot{\bf q}}{|{\bf q}^{\prime}||{\bf q}|}\right], \label{eq:sol-g1-menor}
\end{equation}
and
\begin{equation}
{\cal G}_{>}^{(1)}({\bf q},z,{\bf r^{\prime}})=-e^{-|{\bf q}|z}\int\frac{d^{2}{\bf q}^{\prime}}{\left(2\pi\right)^{2}}p({\bf q},{\bf q}^{\prime},{\bf r^{\prime}})\left[\frac{{\bf q}^{\prime}\cdot{\bf q}}{|{\bf q}^{\prime}||{\bf q}|}+\frac{\epsilon_{1}}{\epsilon_{2}}\right], \label{eq:sol-g1-maior}
\end{equation}
with
\begin{equation}
p({\bf q},{\bf q}^{\prime},{\bf r^{\prime}})=4\pi\frac{\left(\epsilon_{1}-\epsilon_{2}\right)}{\left(\epsilon_{1}+\epsilon_{2}\right)^{2}}h_{(1)}\left({\bf q}-{\bf q}^{\prime}\right)e^{-i{\bf q}^{\prime}\cdot\boldsymbol{r}_{\parallel}^{\prime}}e^{-\left|{\bf q}^{\prime}\right|z^{\prime}}. \label{eq:p}
\end{equation}
Note that, for $\epsilon_{1} \to \infty$ and $\epsilon_{2}=1$, the solution for ${\cal G}^{(1)}({\bf q},z,{\bf r^{\prime}})$ recovers that for a perfectly conducting surface found in Ref. \cite{Clinton-PRB-1985}.

By performing an inverse Fourier transformation on ${\cal G}^{(0)}({\bf q},z,{\bf r^{\prime}})$, one can analytically obtain the solution for $G^{(0)}\left(\textbf{r},\textbf{r}^{\prime}\right)$, which is given by
\begin{equation}
G_{<}^{(0)}\left({\bf r},{\bf r^{\prime}}\right)=\frac{2}{\left(\epsilon_{1}+\epsilon_{2}\right)}\frac{1}{\left|{\bf r}-{\bf r}^{\prime}\right|},
\end{equation}
for $z<0$, and 
\begin{equation}
G_{>}^{(0)}\left({\bf r},{\bf r^{\prime}}\right)=\frac{1}{\epsilon_{2}\left|{\bf r}-{\bf r}^{\prime}\right|}-\frac{1}{\epsilon_{2}}\frac{(\epsilon_{1}-\epsilon_{2})/(\epsilon_{1}+\epsilon_{2})}{\sqrt{|{\bf r}_{\parallel}-{\bf r}_{\parallel}^{\prime}|^{2}+\left(z+z^{\prime}\right)^{2}}}, \label{eq:sol-g0-coord}
\end{equation}
for $z>0$. These solutions are in agreement with those in Ref. \cite{Jackson-Electrodynamics-1998}.
In the same way, one can obtain the solution for $G^{(1)}\left(\textbf{r},\textbf{r}^{\prime}\right)$ performing the inverse Fourier transformation,
\begin{equation}
G^{(1)}\left({\bf r},{\bf r}^{\prime}\right)=\int\frac{d^{2}{\bf q}}{\left(2\pi\right)^{2}}e^{i{\bf q}\cdot{\bf r}_{\parallel}}{\cal G}^{(1)}\left({\bf q},z,{\bf r^{\prime}}_{\parallel},z^{\prime}\right), \label{eq:sol-g1-coord}
\end{equation}
with ${\cal G}^{(1)}$ given by Eqs. \eqref{eq:sol-g1}-\eqref{eq:p}.

It is important to remark that the approximate solution for $G$ [Eq. \eqref{eq:g-coordenadas}, with $\eta=1$] becomes increasingly better as $a/z^\prime\ll 1$.
Besides this, all the perturbative process to obtain Eqs. \eqref{eq:sol-g1}-\eqref{eq:p} is done in terms of the parameter $a/z^\prime$, so that no derivatives of $h$ appear in these equations.
This means that no demand on the smoothness of the interface $z=h({\bf r}_{||})$ is necessary, and thus the approach presented here is valid beyond the PFA \cite{Clinton-PRB-1985,Nogueira-PRA-2021}.

The classical interaction energy between the charge $Q$ and the dielectric $\epsilon_{1}$ depends on the homogeneous part of $G$, namely
\begin{equation}
G_H\left(\textbf{r},\textbf{r}^{\prime}\right)=G\left(\textbf{r},\textbf{r}^{\prime}\right)-\frac{1}{\epsilon_{2}|\textbf{r}-\textbf{r}^{\prime}|}.
\end{equation}
This function $ G_H $ is the solution of the Laplace equation in the presence of the considered surface and, therefore, contains all the information about the surface geometry \cite{Souza-AJP-2013}.
Since the charge is situated at $z^\prime > 0$, only the solutions of $G^{(0)}$ and $G^{(1)}$ for $z>0$ are used.
Thus, from Eqs. \eqref{eq:g-coordenadas} (with $\eta=1$), \eqref{eq:sol-g0-coord} and \eqref{eq:sol-g1-coord}, one can write
\begin{equation}
G_{H}\left(\textbf{r},\textbf{r}^{\prime}\right) \approx -\frac{1}{\epsilon_{2}}\frac{(\epsilon_{1}-\epsilon_{2})/(\epsilon_{1}+\epsilon_{2})}{\sqrt{|{\bf r}_{\parallel}-{\bf r}_{\parallel}^{\prime}|^{2}+\left(z+z^{\prime}\right)^{2}}} + G^{(1)}\left(\textbf{r},\textbf{r}^{\prime}\right). \label{eq:GH}
\end{equation}

The second part of our approach consists of combining  Eq. \eqref{eq:GH} with the method of
Eberlein and Zietal \cite{Eberlein-PRA-2007} to compute the vdW interaction between the polarizable particle and the medium $\epsilon_{1}$.
This method consists of mapping the classical interaction energy $U_\text{cla}$,
\begin{equation}
U_\text{cla}(\textbf{r}_0)=\frac{1}{8\pi\epsilon_{0}}\left.(\textbf{d}\cdot \boldsymbol{ \nabla}^{\prime})(\textbf{d}\cdot\boldsymbol{\nabla})G_{H}(\mathbf{r},\mathbf{r}^{\prime})\right|_{\mathbf{r}=\mathbf{r}^{\prime}=\mathbf{r_{0}}},
\label{eq:Eberlein_Zietalclassic}
\end{equation}
which is the interaction energy for a neutral point particle located at ${\bf r}_{0}=x_0\hat{{\bf x}}+y_0\hat{{\bf y}}+z_0\hat{{\bf z}}$
$(z_0>0)$ and with a dipole moment vector $\textbf{d}$, into the van der Waals interaction $U_{\text{vdW}}$,
\begin{equation}
U_{\text{vdW}}(\mathbf{r}_0)=
\frac{1}{8\pi\epsilon_{0}}\displaystyle\sum_{i,j}\langle \hat{d}_i \hat{d}_j\rangle\boldsymbol{ \nabla}_i\boldsymbol{ \nabla}_j'\left.G_H(\mathbf{r},\mathbf{r'})\right|_{\mathbf{r}=\mathbf{r'}=\mathbf{r}_0},	\label{eq:Eberlein_Zietalquantum}
\end{equation}
where $\hat{d}_i$ $(i,j=\{x,y,z\})$ are the components of the dipole moment operator and
$\langle \hat{d}_i \hat{d}_j\rangle$ is the expectation value of $\hat{d}_i \hat{d}_j$ (see also \cite{Souza-AJP-2013}).

We combine Eq. \eqref{eq:GH} with Eq. 
\eqref{eq:Eberlein_Zietalclassic} (in Sec. \ref{sec-classical-case}), and
with Eq. \eqref{eq:Eberlein_Zietalquantum} (in Sec. \ref{sec-vdw}), 
and investigate their physical implications.

%
\section{Classical Interaction}
\label{sec-classical-case}

Let us consider a neutral polarized particle, with a dipole moment $\textbf{d}$, put at ${\bf r}_{0}=x_0\hat{{\bf x}}+y_0\hat{{\bf y}}+z_0\hat{{\bf z}}$ $(z_0>0)$, in a dielectric $\epsilon_{2}$, as illustrated in Fig. \ref{fig:carga-2-meios-corrug} (with $\textbf{r}^\prime=\textbf{r}_0$). 
The electrostatic interaction $U_\text{cla}$ between this particle and a dielectric $\epsilon_{1}$ [Fig. \ref{fig:carga-2-meios-corrug}] can be written as $U_\text{cla}\approx U^{(0)}_\text{cla} + U^{(1)}_\text{cla}$.
The first term $U^{(0)}_\text{cla}$, which can be written as
\begin{equation}
U_{\text{cla}}^{(0)}(\mathbf{r}_{0})=-\frac{1}{\epsilon_{2}} \left(\frac{\epsilon_{1}-\epsilon_{2}}{\epsilon_{1}+\epsilon_{2}}\right)\frac{d_{x}^{2}+d_{y}^{2}+2d_{z}^{2}}{64\pi\epsilon_{0}z_{0}^{3}},
\label{U-cla-0}
\end{equation}
is the interaction energy between the dipole and the dielectric $\epsilon_{1}$, for the case of a planar interface, as illustrated in Fig. \ref{fig:carga-2-meios-plano}. 
As known, in Eq. \eqref{U-cla-0} when $\epsilon_{2}<\epsilon_{1}$, one has the dipole attracted towards the planar interface, whereas 
if $\epsilon_{2}>\epsilon_{1}$,
it occurs a repulsion and a reduction of the magnitude of the force.
The second term, $ U^{(1)}_\text{cla}$,
which is the first-order correction to $U^{(0)}_\text{cla}$, can be written as
\begin{equation}
U_{\text{cla}}^{(1)}(\mathbf{r}_{0})=\frac{\epsilon_{1}}{\epsilon_{2}}\frac{\left(\epsilon_{1}-\epsilon_{2}\right)}{\left(\epsilon_{1}+\epsilon_{2}\right)^{2}}U_{\text{cond}}^{(1)}(\mathbf{r}_{0})+U_{\text{diel}}^{(1)}(\mathbf{r}_{0}). \label{eq:u1-geral}
\end{equation}
The first term $U_{\text{cond}}^{(1)}$ is the expression obtained for $U_{\text{cla}}^{(1)}$ in Ref. \cite{Nogueira-PRA-2021}, which is related to the interaction between the dipole in vacuum and a grounded conducting corrugated surface, and is given by
\begin{equation}
U_{\text{cond}}^{(1)}(\mathbf{r}_{0})=-\sum_{i,j}\frac{d_{i}d_{j}}{64\pi\epsilon_{0}z_{0}^{4}}\int\frac{d^{2}{\bf q}}{(2\pi)^{2}}\tilde{h}({\bf q})e^{i{\bf q}\cdot{\bf r}_{0\parallel}}\mathcal{I}_{ij}^{\text{cond}}\left(\mathbf{q}z_{0}\right), \label{eq:u1-condutor}
\end{equation}
where $\tilde{h}({\bf q})$ is the Fourier representation of $h(\textbf{r}_{\parallel})$, and the functions $\mathcal{I}_{ij}^{\text{cond}}=\mathcal{I}_{ji}^{\text{cond}}$ are defined as
\begin{align}
\mathcal{I}_{xx}^{\text{cond}}(\mathbf{q}z_{0}) & =\frac{3}{8}q^{2}z_{0}^{2}[qz_{0}K_{3}(qz_{0})-q_{x}^{2}z_{0}^{2}K_{2}(qz_{0})], \nonumber \\
\mathcal{I}_{yy}^{\text{cond}}(\mathbf{q}z_{0}) & =\frac{3}{8}q^{2}z_{0}^{2}[qz_{0}K_{3}(qz_{0})-q_{y}^{2}z_{0}^{2}K_{2}(qz_{0})], \nonumber \\
\mathcal{I}_{zz}^{\text{cond}}(\mathbf{q}z_{0}) & =[2q^{2}z_{0}^{2}+\frac{3}{8}q^{4}z_{0}^{4}]K_{2}(qz_{0})+\frac{1}{4}q^{3}z_{0}^{3}K_{3}(qz_{0}), \nonumber \\
\mathcal{I}_{xy}^{\text{cond}}(\mathbf{q}z_{0}) & =-\frac{3}{8}q_{x}q_{y}q^{2}z_{0}^{4}K_{2}(qz_{0}),  \label{eq:Iij-condutor}\\
\mathcal{I}_{xz}^{\text{cond}}(\mathbf{q}z_{0}) & =iq_{x}q^{2}z_{0}^{3}[K_{2}(qz_{0})-\frac{3}{8}qz_{0}K_{3}(qz_{0})],\nonumber \\
\mathcal{I}_{yz}^{\text{cond}}(\mathbf{q}z_{0}) & =iq_{y}q^{2}z_{0}^{3}[K_{2}(qz_{0})-\frac{3}{8}qz_{0}K_{3}(qz_{0})],\nonumber 
\end{align}
where $K_{2}$ and $K_{3}$ are modified Bessel functions of the second kind.
The second term, $U_{\text{diel}}^{(1)}$, in Eq. \eqref{eq:u1-geral}, is a correction in the interaction energy which arises specifically from the consideration of a dielectric as a corrugated surface.
We also write $U_{\text{diel}}^{(1)}$ in terms of $\tilde{h}({\bf q})$, so that, it is given by
\begin{align}
U_{\text{diel}}^{(1)}(\mathbf{r}_{0}) & =-\frac{(\epsilon_{1}-\epsilon_{2})}{(\epsilon_{1}+\epsilon_{2})^{2}}\sum_{i,j}\frac{d_{i}d_{j}}{64\pi\epsilon_{0} z_{0}^{4}}\int\frac{d^{2}\mathbf{q}}{(2\pi)^{2}}\tilde{h}(\mathbf{q})\nonumber \\
& \times e^{i\mathbf{q}\cdot{\bf r}_{0\parallel}}\mathcal{I}_{ij}^{\text{diel}}(\mathbf{q}z_{0}),
\label{eq:u1-dieletrico} 
\end{align}
where the functions $\mathcal{I}_{ij}^{\text{diel}}=\mathcal{I}_{ji}^{\text{diel}}$ are defined as
\begin{align}
\mathcal{I}_{xx}^{\text{diel}}(\mathbf{q}z_{0})= & (4q_{x}^{2}z_{0}^{2}+3q^{2}z_{0}^{2}+\frac{3}{8}q_{x}^{2}q^{2}z_{0}^{4})K_{2}(qz_{0}) \nonumber \\
& -(q_{x}^{2}qz_{0}^{3}+\frac{3}{8}q^{3}z_{0}^{3})K_{3}(qz_{0}), \nonumber \\
\mathcal{I}_{yy}^{\text{diel}}(\mathbf{q}z_{0})= & (4q_{y}^{2}z_{0}^{2}+3q^{2}z_{0}^{2}+\frac{3}{8}q_{y}^{2}q^{2}z_{0}^{4})K_{2}(qz_{0}) \nonumber \\
& -(q_{y}^{2}qz_{0}^{3}+\frac{3}{8}q^{3}z_{0}^{3})K_{3}(qz_{0}), \label{eq:Iij-dieletrico}\\
\mathcal{I}_{zz}^{\text{diel}}(\mathbf{q}z_{0})= & \frac{3}{4}q^{3}z_{0}^{3}K_{3}(qz_{0})-\frac{3}{8}q^{4}z_{0}^{4}K_{2}(qz_{0}), \nonumber \\
\mathcal{I}_{xy}^{\text{diel}}(\mathbf{q}z_{0})= & q_{x}q_{y}z_{0}^{2}[(4+\frac{3}{8}q^{2}z_{0}^{2})K_{2}(qz_{0})-qz_{0}K_{3}(qz_{0})], \nonumber \\
\mathcal{I}_{xz}^{\text{diel}}(\mathbf{q}z_{0})= & iq_{x}q^{2}z_{0}^{3}[\frac{3}{8}qz_{0}K_{3}(qz_{0})-2K_{2}(qz_{0})], \nonumber \\
\mathcal{I}_{yz}^{\text{diel}}(\mathbf{q}z_{0})= & iq_{y}q^{2}z_{0}^{3}[\frac{3}{8}qz_{0}K_{3}(qz_{0})-2K_{2}(qz_{0})]. \nonumber  
\end{align} 
Substituting Eqs. \eqref{eq:u1-condutor} and \eqref{eq:u1-dieletrico} in Eq. \eqref{eq:u1-geral}, we write $U_{\text{cla}}^{(1)}$ as 
\begin{align}
U_{\text{cla}}^{(1)}(\mathbf{r}_{0})= & -\frac{\epsilon_{1}^2(1-\frac{\epsilon_{2}}{\epsilon_{1}})}{\epsilon_{2}\left(\epsilon_{1}+\epsilon_{2}\right)^{2}}\sum_{i,j}\frac{d_{i}d_{j}}{64\pi\epsilon_{0}z_{0}^{4}}\int\frac{d^{2}\mathbf{q}}{\left(2\pi\right)^{2}}\tilde{h}\left(\mathbf{q}\right) \nonumber \\
& \times e^{i\mathbf{q}\cdot{\bf r}_{0\parallel}} \left[\mathcal{I}_{ij}^{\text{cond}}\left(\mathbf{q}z_{0}\right)+\frac{\epsilon_{2}}{\epsilon_{1}}\mathcal{I}_{ij}^{\text{diel}}\left(\mathbf{q}z_{0}\right)\right]. \label{eq:principal}
\end{align}
Making $\epsilon_{1}\to \infty$ and $\epsilon_{2} = 1$ in Eq. \eqref{eq:principal}, 
one recovers $U_{\text{cla}}^{(1)}=U_{\text{cond}}^{(1)}$, obtained in Ref. \cite{Nogueira-PRA-2021}.

Let us investigate the case of a sinusoidal corrugated surface with amplitude $a$ and corrugation period $\lambda$, which is described by $h(x)=a\cos(k x)$, where  $k = 2\pi/\lambda$ and $a\ll z_0$.
Then, the integrals in Eq. \eqref{eq:principal} can be analytically solved \cite{Gradshteyn-Table-2007, Watson-BesselFunctions-1944}, resulting in
\begin{align}
U_{\text{cla}}^{(1)}(\mathbf{r}_{0})=-\frac{\epsilon_{1}^2 }{\epsilon_{2}(\epsilon_{1}+\epsilon_{2})^{2}}& \frac{3aA(\frac{\epsilon_{2}}{\epsilon_{1}},d_{i}d_{j},kz_{0})}{ 512\pi\epsilon_{0}z_{0}^{4}} \nonumber \\ 
\times   \cos&\left[kx_{0}-\delta\left(\frac{\epsilon_{2}}{\epsilon_{1}},d_{i}d_{j},kz_{0}\right)\right],
\label{eq:u1-cos}
\end{align}
where $\delta$ is  a nontrivial phase function defined by
\begin{equation}
	\sin(\delta)=\frac{B(\frac{\epsilon_{2}}{\epsilon_{1}},d_i d_j,kz_{0})}{A(\frac{\epsilon_{2}}{\epsilon_{1}},d_i d_j,kz_{0})},\;\;\;\cos(\delta)=\frac{C(\frac{\epsilon_{2}}{\epsilon_{1}},d_i d_j,kz_{0})}{A(\frac{\epsilon_{2}}{\epsilon_{1}},d_i d_j,kz_{0})}, 
	\label{eq:delta}
\end{equation}
with $A=\sqrt{B^{2}+C^{2}}$,
\begin{align}
B\left(\frac{\epsilon_{2}}{\epsilon_{1}},d_{i}d_{j},kz_{0}\right) = &-2d_{x}d_{z}\left(1-\frac{\epsilon_{2}}{\epsilon_{1}}\right) \nonumber \\
&\times[\mathcal{R}_{xz}^{\text{cond}}(kz_{0})+\frac{\epsilon_{2}}{\epsilon_{1}}\mathcal{R}_{xz}^{\text{diel}}(kz_{0})], \label{eq:B} \\
C\left(\frac{\epsilon_{2}}{\epsilon_{1}},d_{i}d_{j},kz_{0}\right) = & \left(1-\frac{\epsilon_{2}}{\epsilon_{1}}\right)\sum_{i}d_{i}^{2} \nonumber \\
&\times [\mathcal{R}_{ii}^{\text{cond}}\left(kz_{0}\right)+\frac{\epsilon_{2}}{\epsilon_{1}}\mathcal{R}_{ii}^{\text{diel}}\left(kz_{0}\right)], \label{eq:C} 
\end{align}
the functions ${\cal R}_{ij}^{\text{cond}}(kz_{0})$ are defined by:
\begin{align}
\mathcal{R}_{xx}^{\text{cond}}\left(u\right) & =u^{3}K_{3}\left(u\right)-u^{4}K_{2}\left(u\right); \nonumber \\
\mathcal{R}_{yy}^{\text{cond}}\left(u\right) & =u^{3}K_{3}\left(u\right);  \label{eq:rij-condutor} \\
\mathcal{R}_{zz}^{\text{cond}}\left(u\right) & =\left(\frac{16}{3}u^{2}+u^{4}\right)K_{2}\left(u\right)+\frac{2}{3}u^{3}K_{3}\left(u\right); \nonumber \\
\mathcal{R}_{xz}^{\text{cond}}\left(u\right) & =\frac{8}{3}u^{3}K_{2}\left(u\right)-u^{4}K_{3}\left(u\right), \nonumber
\end{align}
and the functions ${\cal R}_{ij}^{\text{diel}}(kz_{0})$ are defined by:
\begin{align}
\mathcal{R}_{xx}^{\text{diel}}\left(u\right) & =\left(\frac{56}{3}u^{2}+u^{4}\right)K_{2}\left(u\right)-\frac{11}{3}u^{3}K_{3}\left(u\right); \nonumber \\
\mathcal{R}_{yy}^{\text{diel}}\left(u\right) & =8u^{2}K_{2}\left(u\right)-u^{3}K_{3}\left(u\right);  \label{eq:rij-dieletrico} \\
\mathcal{R}_{zz}^{\text{diel}}\left(u\right) & =2u^{3}K_{3}\left(u\right)-u^{4}K_{2}\left(u\right); \nonumber \\
\mathcal{R}_{xz}^{\text{diel}}\left(u\right) & =u^{4}K_{3}\left(u\right)-\frac{16}{3}u^{3}K_{2}\left(u\right). \nonumber 
\end{align}

\begin{figure}[h]
\centering
\epsfig{file=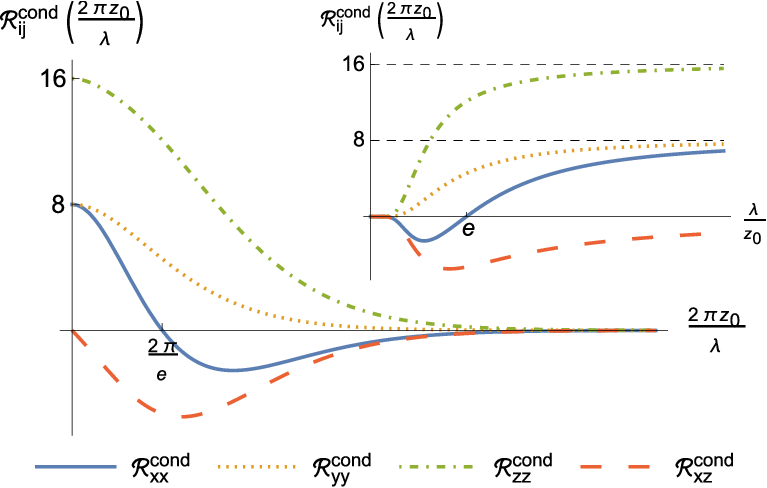,  width=0.9 \linewidth}
\caption{
The behavior of ${\cal R}_{ij}^{\text{cond}}(2\pi z_0/\lambda)$ versus $2\pi z_0/\lambda$.
The function $\mathcal{R}_{xx}^{\text{cond}}$ (solid line) changes its sign at $2\pi z_0/\lambda \approx 2\pi/e$, where $e$ is the Euler number.
In the inset we also show the behavior of ${\cal R}_{ij}^{\text{cond}}(2\pi z_0/\lambda)$ versus $\lambda/z_0$.
Note that the function $\mathcal{R}_{xx}^{\text{cond}}$ changes its sign at $\lambda/z_0 \approx e$.
}
\label{fig:rij-condutor}
\end{figure}

\begin{figure}[h]
\centering
\epsfig{file=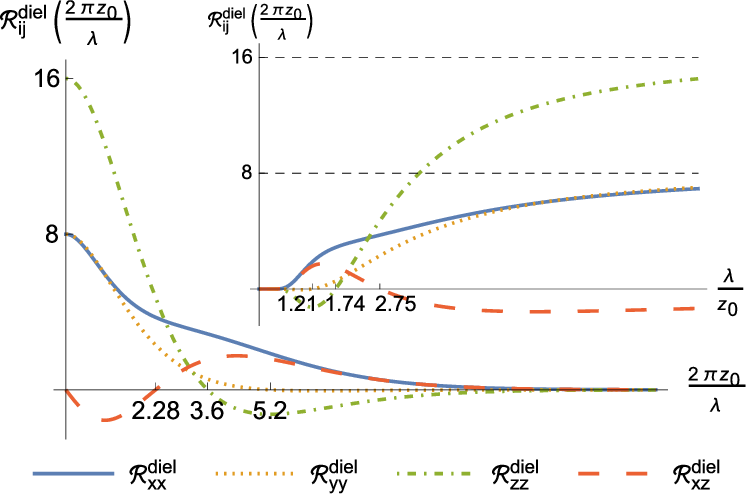,  width=1 \linewidth}
\caption{
The behavior of ${\cal R}_{ij}^{\text{diel}}(2\pi z_0/\lambda)$ versus $2\pi z_0/\lambda$.
The functions $\mathcal{R}_{yy}^{\text{diel}}$ (dotted line), $\mathcal{R}_{zz}^{\text{diel}}$ (dot-dashed line) and $\mathcal{R}_{xz}^{\text{diel}}$ (dashed line) change their sign at $2\pi z_0/\lambda \approx 5.2$, $2\pi z_0/\lambda \approx 3.6$ and $2\pi z_0/\lambda \approx 2.28$, respectively.
In the inset we also show the behavior of ${\cal R}_{ij}^{\text{diel}}(2\pi z_0/\lambda)$ versus $\lambda/z_0$.
Note that the function $\mathcal{R}_{xx}^{\text{diel}}$ is the only one that not changes its sign.
}
\label{fig:rij-dieletrico}
\end{figure}

We highlight that, among the functions ${\cal R}_{ij}^{\text{cond}}$ in Eqs. \eqref{eq:rij-condutor}, just ${\cal R}_{xx}^{\text{cond}}$ changes its sign, which happens at the point $kz_0\approx 2\pi/e$, with this value corresponding to $\lambda/z_0\approx e$, where $e$ is the Euler number (see Fig. \ref{fig:rij-condutor}).
In contrast, among the functions ${\cal R}_{ij}^{\text{diel}}$ in Eqs. \eqref{eq:rij-dieletrico}, just ${\cal R}_{xx}^{\text{diel}}$ does not changes its sign. 
The functions ${\cal R}_{yy}^{\text{diel}}$, ${\cal R}_{zz}^{\text{diel}}$ and ${\cal R}_{xz}^{\text{diel}}$ change their signs at the points $kz_0\approx 5.2$, $kz_0\approx 3.6$ and $kz_0\approx 2.28$, respectively, with these values corresponding to $\lambda/z_0\approx 1.21$, $\lambda/z_0\approx 1.74$ and $\lambda/z_0\approx 2.75$, respectively (see Fig. \ref{fig:rij-dieletrico}).
These behaviors for the functions ${\cal R}_{ij}^{\text{cond}}$ and ${\cal R}_{ij}^{\text{diel}}$ affect the behavior of the phase function $\delta$ in a nontrivial way, as we will see soon.
In order to simplify the discussions, hereafter we make our analysis in terms of the parameter $\lambda/z_0$.

\begin{figure}[h]
\centering 
\epsfig{file=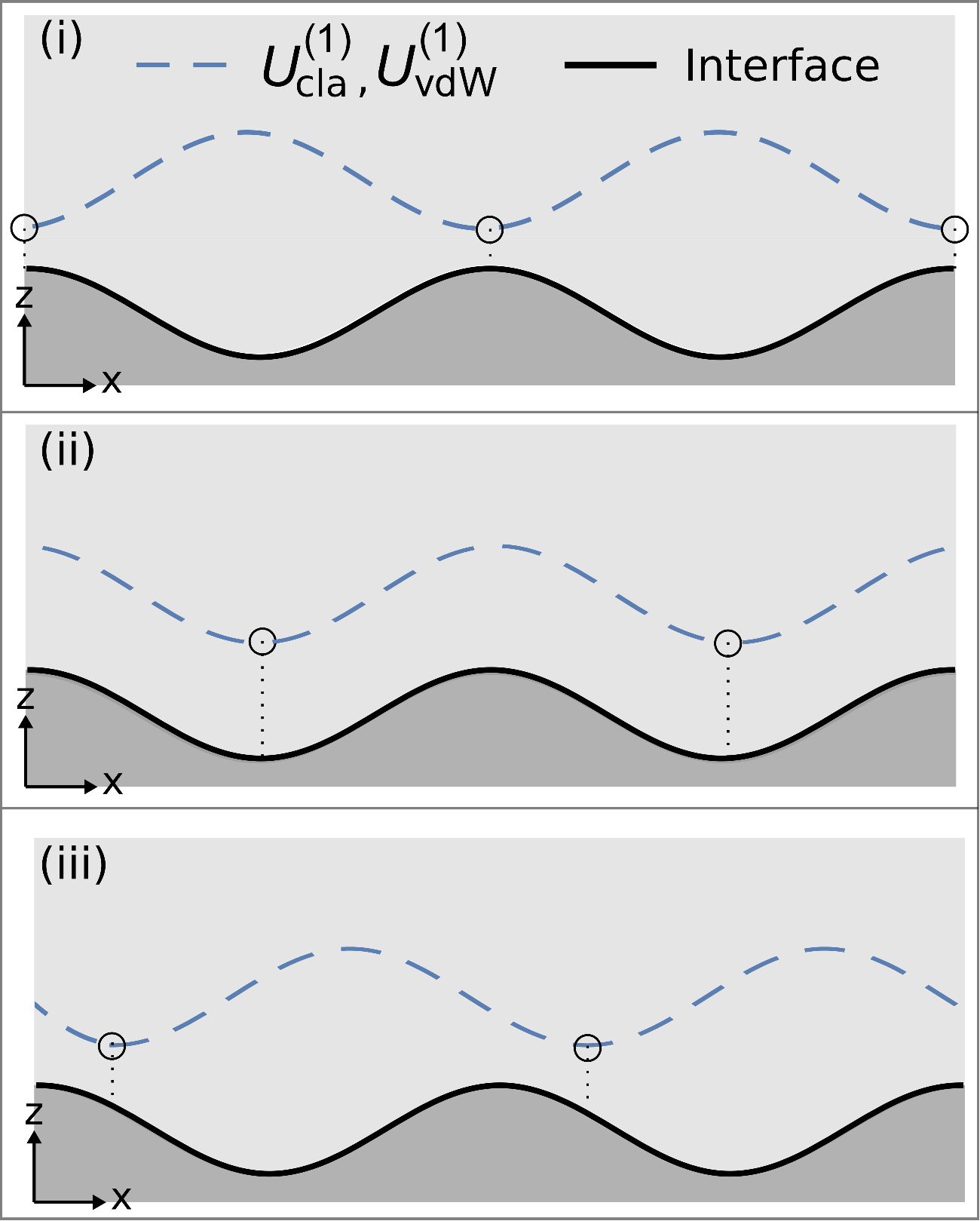,  width=0.7 \linewidth}  
\caption{
Illustration of the stable equilibrium points (indicated by the circles) of $U^{(1)}_\text{cla}$ [Eq. \eqref{eq:u1-cos}] (dashed lines). 
Note that these equilibrium points can be over the corrugation peaks [peak regime, (i)], valleys [valley regime, (ii)], or over points between a peak and a valley [intermediate regime, (iii)]. 
The same figure also illustrates the stable equilibrium points of the vdW energy $U^{(1)}_\text{vdW}$ [Eq. \eqref{eq:potential-energy-CP}]. 
}
\label{fig:regimes}
\end{figure}

Considering the particle fixed at $z=z_0$, 
the stable equilibrium points of $U^{(1)}_\text{cla}$
in Eq. \eqref{eq:u1-cos} depend on 
the phase $\delta$, so that the points can be over the corrugation peaks, valleys, or over points between a peak and a valley, with such behaviors named in Ref. \cite{Nogueira-PRA-2021} as peak, valley and intermediate regimes, respectively (see Fig. \ref{fig:regimes}).

Let us consider the components of the dipole vector in spherical coordinates, thus $d_x = |\textbf{d}| \sin\theta \cos\phi $, $d_y = |\textbf{d}| \sin\theta \sin\phi $ and $d_z = |\textbf{d}| \cos\theta $.
As in Ref. \cite{Nogueira-PRA-2021}, we have peak or valley regimes when $B=0$, which occurs when $\phi=\pi/2$ (dipole oriented in the $zy$-plane), $\theta = 0$ (dipole oriented in $z$-direction) or $\theta = \pi/2$ (dipole oriented in the $xy$-plane). 
When $B=0$ occurs, the sign of the function $C$ determines if we have peak [Fig. \ref{fig:regimes}(i)] or valley regime [Fig. \ref{fig:regimes}(ii)], specifically, $C>0$ results in $\delta = 0$ (peak regime), whereas, $C<0$ results in $\delta = \pi$ (valley regime).
When $B \neq 0$, which occurs when $\theta \neq 0, \pi/2$ (with $\phi \neq \pi/2$), we have the intermediate regime [Fig. \ref{fig:regimes}(iii)].
In this case, the values $x_{\text{min}}$ (minimum values of $U^{(1)}_\text{cla}$) can be analytically computed by solving the equation
\begin{equation}
kx_{\text{min}}-\delta(d_i d_j,kz_{0}) = 2n\pi, \label{eq:interm}
\end{equation}
with $ n \in \mathbb{Z} $, and the phase function $\delta$ obtained from Eqs. \eqref{eq:delta}.

\begin{figure}[h]
\centering  
\subfigure[]{\label{fig:regiao-epsilon0}\epsfig{file=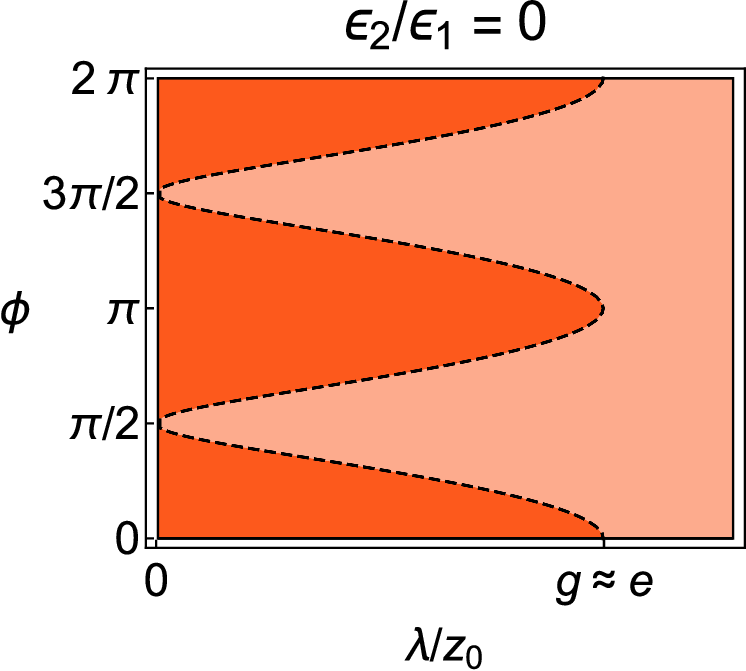, width=0.25 \linewidth}}
\hspace{6mm}
\subfigure[]{\label{fig:regiao-epsilon05}\epsfig{file=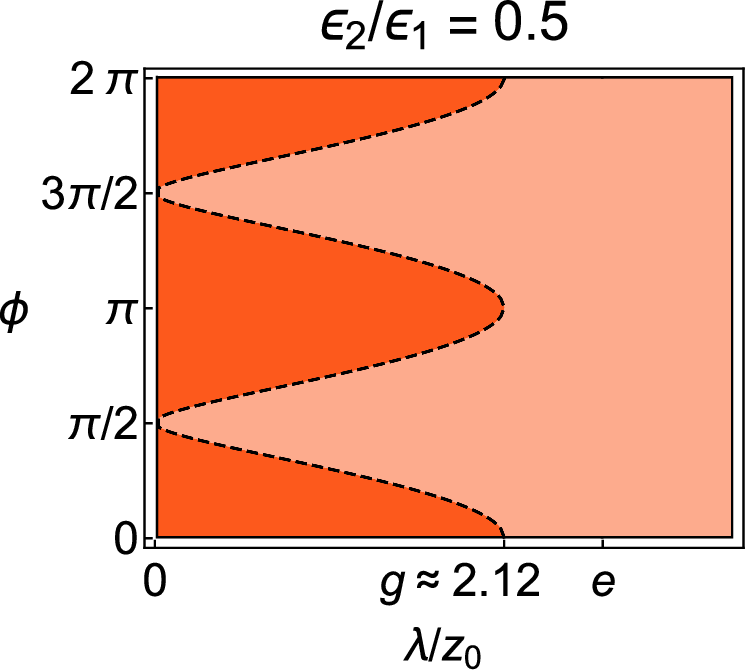, width=0.25 \linewidth}}
\hspace{6mm}
\subfigure[]{\label{fig:regiao-epsilon099}\epsfig{file=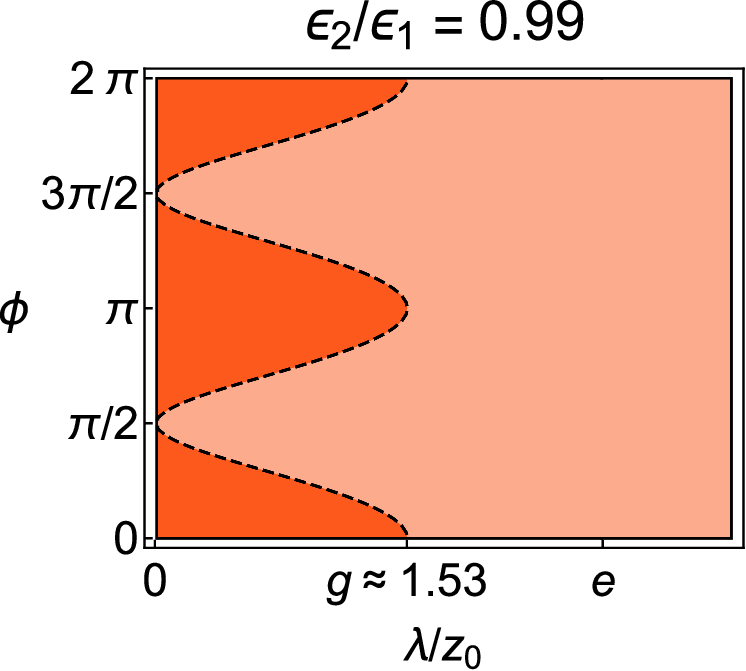, width=0.25 \linewidth}}
\vspace{1mm}
\subfigure[]{\label{fig:regiao-epsilon101}\epsfig{file=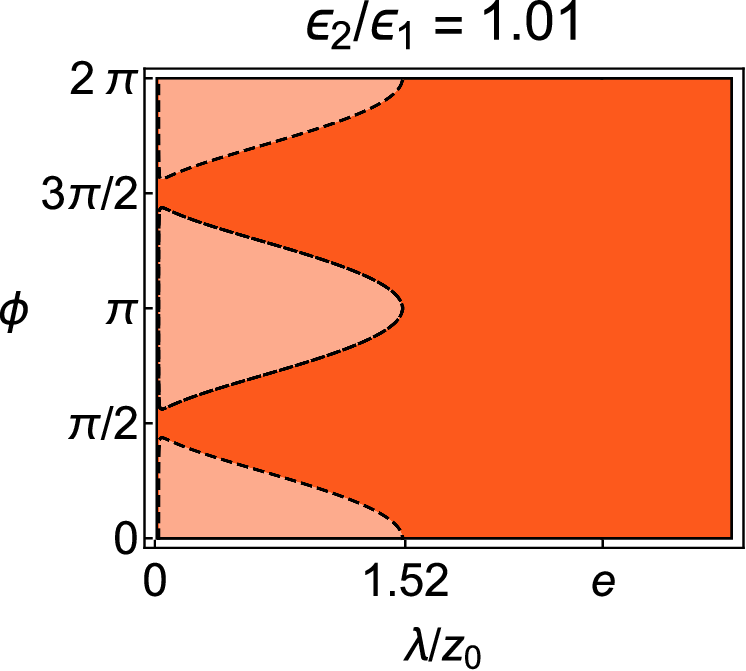, width=0.25 \linewidth}}
\hspace{6mm}
\subfigure[]{\label{fig:regiao-epsilon11}\epsfig{file=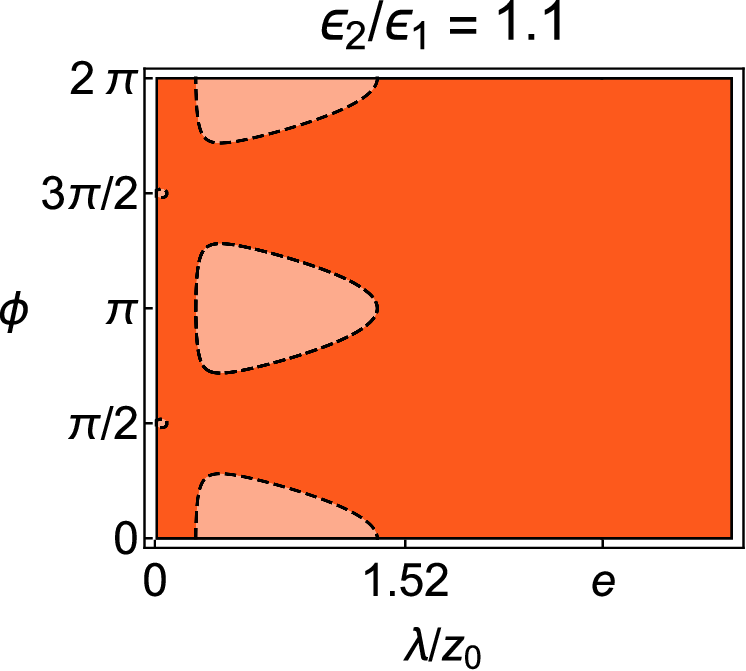, width=0.25 \linewidth}}
\hspace{6mm}
\subfigure[]{\label{fig:regiao-epsilon12}\epsfig{file=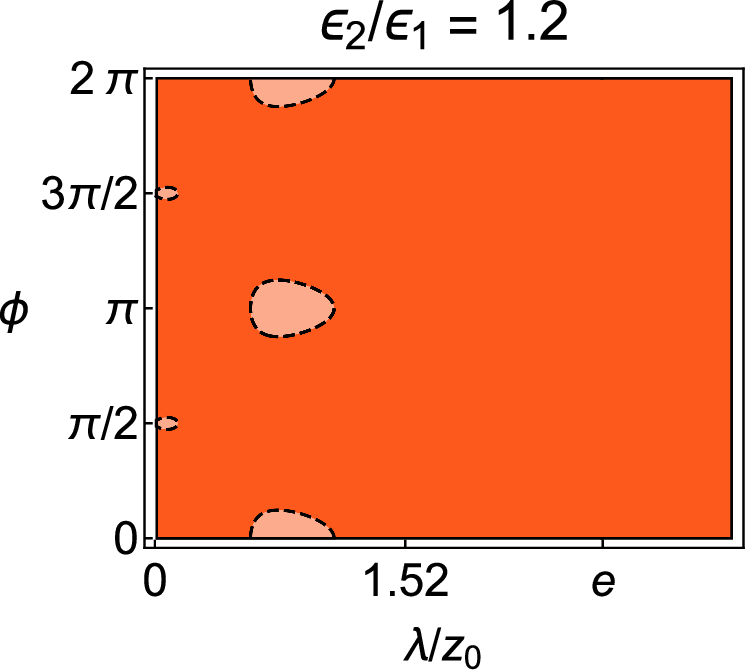, width=0.25 \linewidth}}
\vspace{1mm}
\subfigure[]{\label{fig:regiao-epsilon13}\epsfig{file=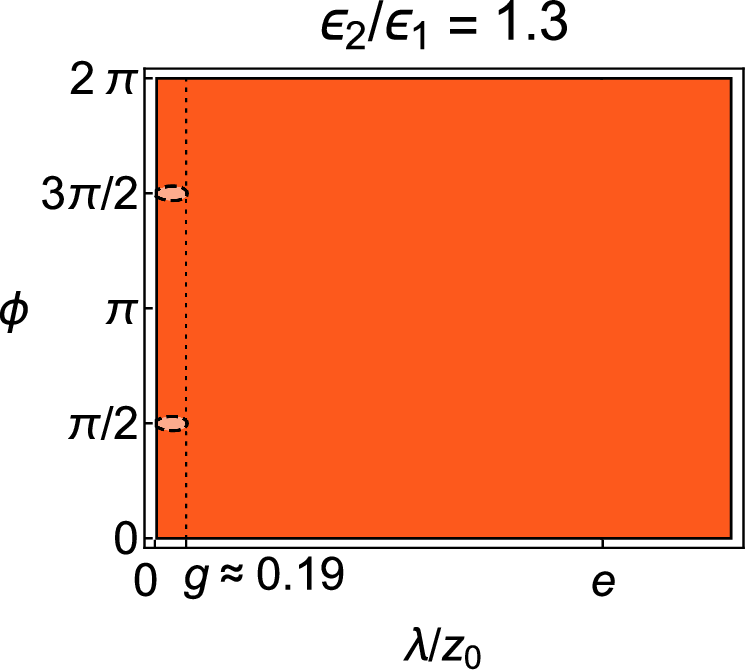, width=0.25 \linewidth}}
\hspace{6mm}
\subfigure[]{\label{fig:regiao-epsilon5}\epsfig{file=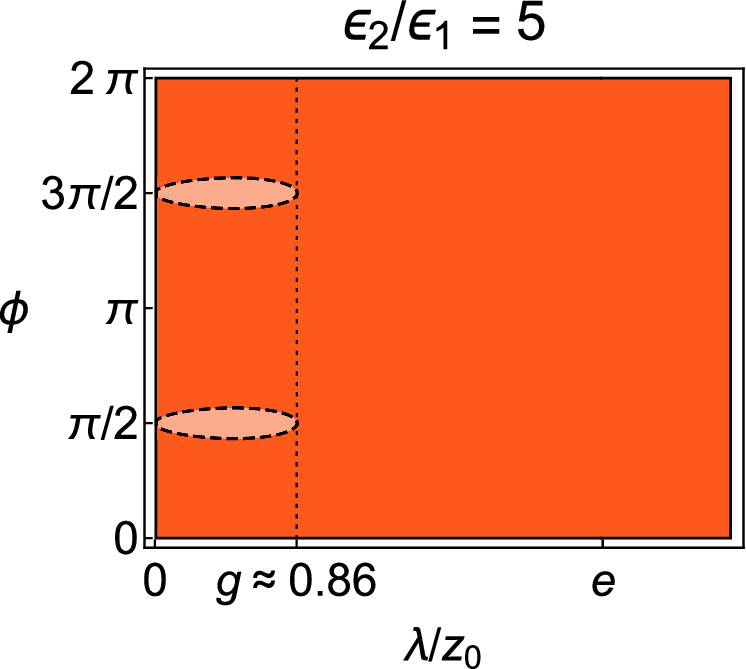, width=0.25 \linewidth}}
\hspace{6mm}
\subfigure[]{\label{fig:regiao-epsilon100}\epsfig{file=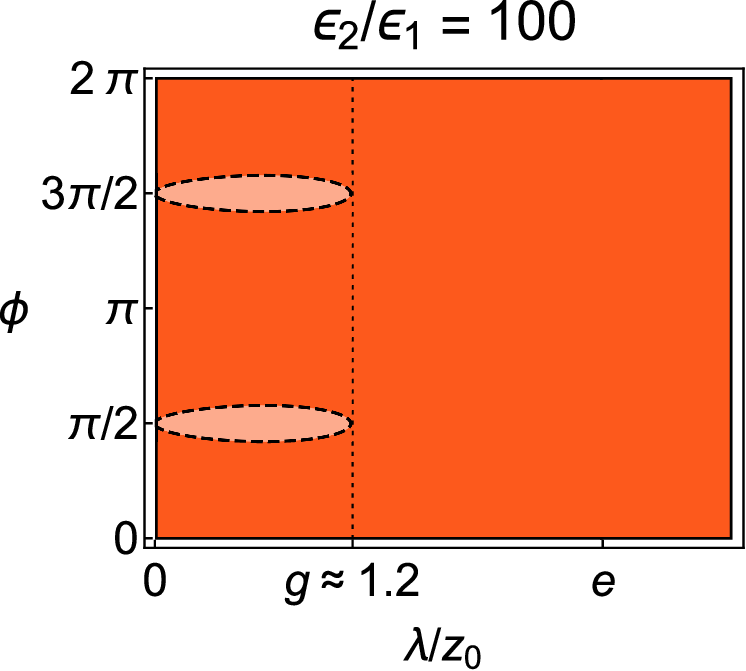, width=0.25 \linewidth}}
\caption{ 
Each figure, representing a configuration space $\phi$ (vertical axis) versus $\lambda/z_0$ (horizontal axis), for a given value of ${\epsilon_{2}}/{\epsilon_{1}}$, show the behavior of $x_{\text{min}}$ of $U^{(1)}_\text{cla}$ for a particle with dipole moment $\textbf{d}$ in the $xy$ plane $(\theta=\pi/2)$. 
The dark regions represent $C(\theta=\pi/2)<0$, and correspond to the valley regime.
The lighter regions represent $C(\theta=\pi/2) > 0$, and correspond to the peak regime.
The border between the lighter and dark regions (dashed lines) corresponds to the situations where $C(\theta=\pi/2) = 0$, which means that $ U_{\text{cla}}^{(1)} = 0 $ and that the lateral force vanishes.
The considered values for ${\epsilon_{2}}/{\epsilon_{1}}$ are: (a) ${\epsilon_{2}}/{\epsilon_{1}} \to 0$ (the particular case addressed in Ref. \cite{Nogueira-PRA-2021}), (b) ${\epsilon_{2}}/{\epsilon_{1}} = 0.5$, (c) ${\epsilon_{2}}/{\epsilon_{1}} = 0.99$, (d) ${\epsilon_{2}}/{\epsilon_{1}} = 1.01$, (e) ${\epsilon_{2}}/{\epsilon_{1}} = 1.1$, (f) ${\epsilon_{2}}/{\epsilon_{1}} = 1.2$, (g) ${\epsilon_{2}}/{\epsilon_{1}} = 1.3$, (h) ${\epsilon_{2}}/{\epsilon_{1}} = 5$ and (i) ${\epsilon_{2}}/{\epsilon_{1}} = 100$.
Note that Figs. (a), (b), and (c) have in common that ${\epsilon_{2}}/{\epsilon_{1}<1}$, whereas
from (d) to (i), we have ${\epsilon_{2}}/{\epsilon_{1}>1}$.
}
\label{fig:regiao-epsilon}
\end{figure}

In this paper, the peak, valley and intermediate regimes are strongly affected by the values of ${\epsilon_{1}}$ and ${\epsilon_{2}}$, since they can affect the sign of the functions $B$ and $C$.
When the dipole is oriented in the $xy$-plane, we can have peak or valley regimes. 
The dependence of these regimes on $\phi$ and $\lambda/z_0$, for several values of ${\epsilon_{2}}/{\epsilon_{1}}$, is shown in Fig. \ref{fig:regiao-epsilon}.
In all these figures, the occurrence of peak and valley regimes are illustrated by the lighter and dark regions, respectively.
The border between these two regions (dashed lines in Fig. \ref{fig:regiao-epsilon}) corresponds to the situations where $U^{(1)}_\text{cla}=0$, which means that the lateral force vanishes.

When ${\epsilon_{2}}/{\epsilon_{1}} < 1 $, we have Fig. \ref{fig:regiao-epsilon0} representing the case of a conducting corrugated surface investigated in Ref. \cite{Nogueira-PRA-2021}, and Figs. \ref{fig:regiao-epsilon05} and \ref{fig:regiao-epsilon099} representing the cases where ${\epsilon_{2}}/{\epsilon_{1}} = 0.5$ and ${\epsilon_{2}}/{\epsilon_{1}} = 0.99$, respectively.
Note that the behaviors illustrated in these figures are similar to each other, with a reduction in the dark region as ${\epsilon_{2}}/{\epsilon_{1}}$ increases,
but we remark that in the limit ${\epsilon_{2}}/{\epsilon_{1}}\to 1^-$ the dark region does not 
vanish, and is limited at $\lambda/z_0=g\approx 1.52$.
By comparing Fig. \ref{fig:regiao-epsilon099} with Fig. \ref{fig:regiao-epsilon101} (${\epsilon_{2}}/{\epsilon_{1}} = 0.99 $ and ${\epsilon_{2}}/{\epsilon_{1}} = 1.01 $, respectively), we can clearly see an inversion $\text{lighter}\leftrightarrow\text{dark}$, as expected.
However, other remarkable behavior is that the lighter region 
in Fig. \ref{fig:regiao-epsilon101} does not start from zero;
in other words, in the  limit ${\epsilon_{2}}/{\epsilon_{1}}\to 1^+$
the lighter region does not vanish, 
and is limited at $\lambda/z_0\approx 1.52$.
Following with our analysis of the transition from Fig. \ref{fig:regiao-epsilon099} to \ref{fig:regiao-epsilon101},
we observe that the inversion from lighter to dark (and vice-versa) is not exact,
but there is a subtle change in the shape of the Fig. \ref{fig:regiao-epsilon101}, 
which is another manifestation of the non-triviality of the results.
This nontrivial behavior becomes more evident as ${\epsilon_{2}}/{\epsilon_{1}}$ increases, as we can see 
from Fig. \ref{fig:regiao-epsilon101} to \ref{fig:regiao-epsilon11} (${\epsilon_{2}}/{\epsilon_{1}} = 1.01 $ and ${\epsilon_{2}}/{\epsilon_{1}} = 1.1 $, respectively).
Note that Fig. \ref{fig:regiao-epsilon11} cannot be obtained by a 
mere inversion $\text{lighter}\leftrightarrow\text{dark}$ 
(or $\text{peak}\leftrightarrow\text{valley}$) in Figs.  	
\ref{fig:regiao-epsilon0}, \ref{fig:regiao-epsilon05} and \ref{fig:regiao-epsilon099}
(for which ${\epsilon_{2}}/{\epsilon_{1}} <1$).
%
From Fig. \ref{fig:regiao-epsilon11} to \ref{fig:regiao-epsilon12} (${\epsilon_{2}}/{\epsilon_{1}} = 1.1 $ and ${\epsilon_{2}}/{\epsilon_{1}} = 1.2 $, respectively),
one can see a reduction of the areas related to the peak regime, with its possibilities
occurring for the particle oriented around $\phi=0,\pi/2,\pi,3\pi/2$.  
In Figs. \ref{fig:regiao-epsilon13}, 
\ref{fig:regiao-epsilon5} and \ref{fig:regiao-epsilon100}
(${\epsilon_{2}}/{\epsilon_{1}} = 1.3,{\epsilon_{2}}/{\epsilon_{1}} = 5 $, and ${\epsilon_{2}}/{\epsilon_{1}} = 100 $, respectively), we have the elimination
of the possibilities of peak regime around $\phi=0,\pi$, still remaining only
around $\phi=\pi/2,3\pi/2$.

\begin{figure}[h]
	\centering  
	\subfigure[]{\label{fig:regiao-x}\epsfig{file=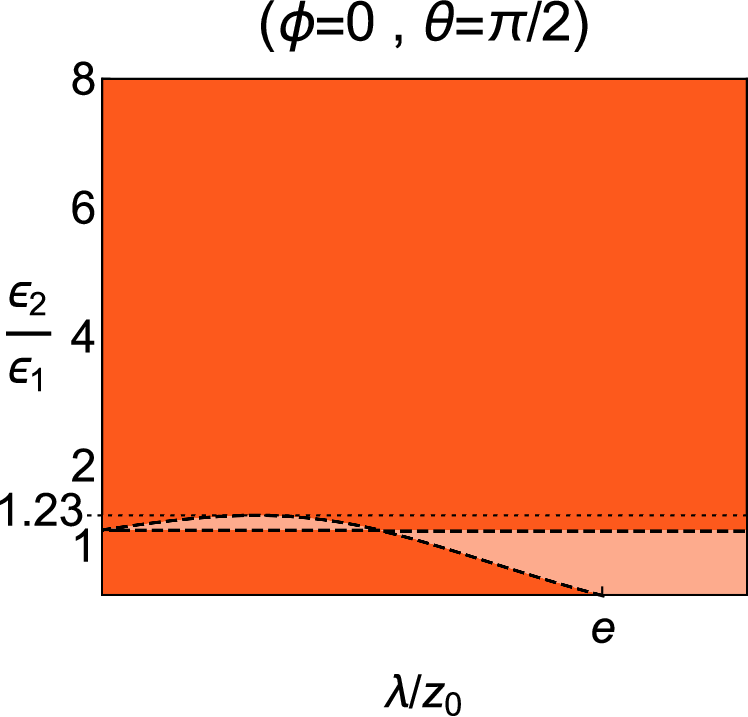, width=0.25 \linewidth}}
	\hspace{6mm}
	\subfigure[]{\label{fig:regiao-y}\epsfig{file=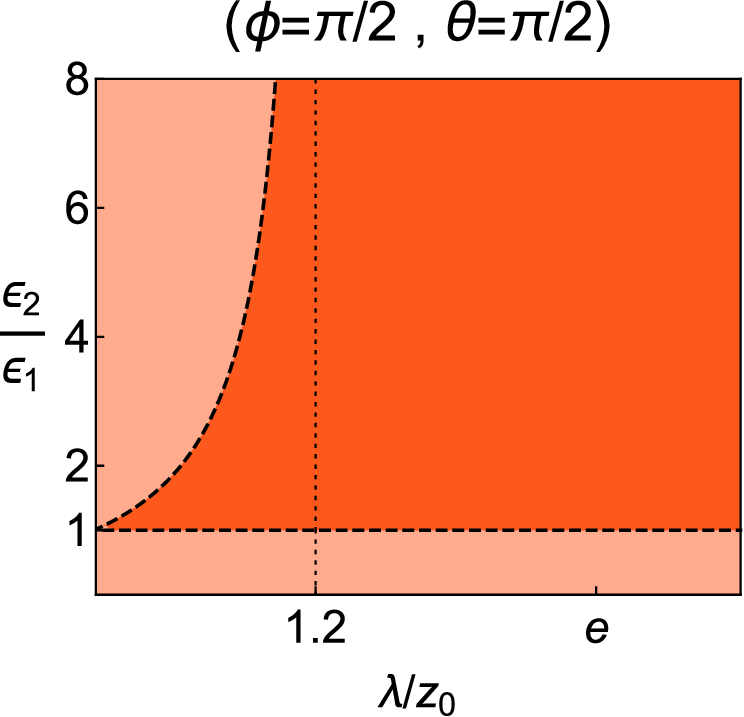, width=0.25 \linewidth}}
	\hspace{6mm}
	\subfigure[]{\label{fig:regiao-z}\epsfig{file=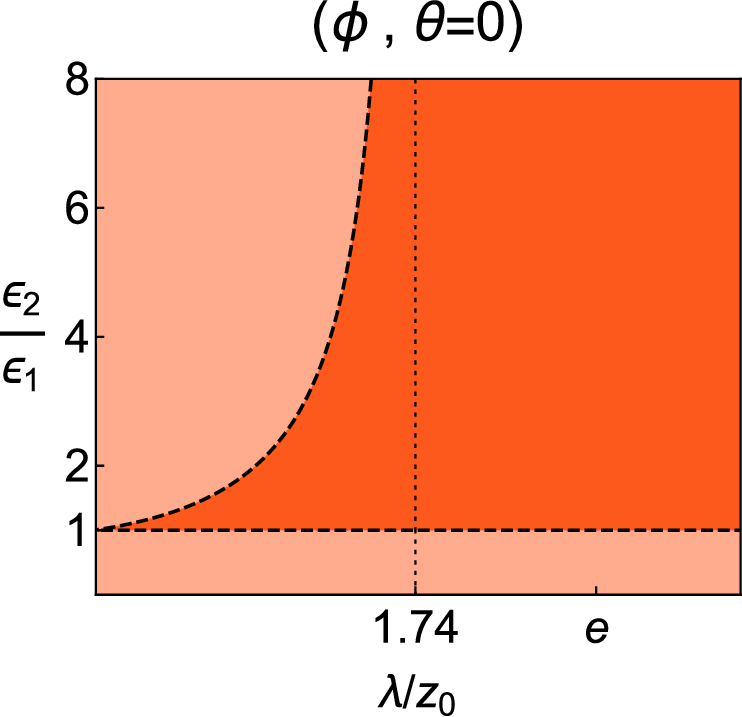, width=0.25 \linewidth}}
	\caption{
	Each figure, representing a configuration space ${\epsilon_{2}}/{\epsilon_{1}}$ (vertical axis) versus $\lambda/z_0$ (horizontal axis), show the behavior of $x_{\text{min}}$ of $U^{(1)}_\text{cla}$ for a particle with dipole moment $\textbf{d}$ oriented in: (a) $x$-direction $(\phi=0$ and $\theta = \pi/2)$; (b) $y$-direction $(\phi=\pi/2$ and $\theta = \pi/2)$; (c) $z$-direction $(\theta = 0)$.
	The dark regions represent $C(\theta=\pi/2)<0$, and correspond to the valley regime.
	The lighter regions represent $C(\theta=\pi/2) > 0$, and correspond to the peak regime.
	}
	\label{fig:regiao-xyz}
\end{figure}

Another way to visualize the nontrivial behavior shown in Figs. \ref{fig:regiao-epsilon0} - \ref{fig:regiao-epsilon100} is focusing on the evolution of the peak and valley regimes only for the particle oriented with $\phi=0,\pi$ (the dipole oriented in $x$-direction), or $\phi=\pi/2,3\pi/2$ (the dipole oriented in $y$-direction).
This evolution, shown in Figs. \ref{fig:regiao-epsilon0} - \ref{fig:regiao-epsilon100},
for specific values of ${\epsilon_{2}}/{\epsilon_{1}}$, is 
described, in Figs. \ref{fig:regiao-x} and \ref{fig:regiao-y}, for continuous values of ${\epsilon_{2}}/{\epsilon_{1}}$. 
In Figs. \ref{fig:regiao-x} and \ref{fig:regiao-y}, note that, for ${\epsilon_{2}}/{\epsilon_{1}} < 1$, we have transition between the peak and valley regimes only when the dipole is oriented in $x$-direction
$(\phi=0,\pi)$, as we can also see from Figs. \ref{fig:regiao-epsilon0} - \ref{fig:regiao-epsilon099}.
Otherwise, for ${\epsilon_{2}}/{\epsilon_{1}} > 1$, we can have transition between the peak and valley regimes when the dipole is oriented in both $x$- or $y$-direction.
In Fig. \ref{fig:regiao-x} (dipole oriented in $x$-direction), we highlight the value ${\epsilon_{2}}/{\epsilon_{1}}\approx 1.23 $, above which we only have valley regime,
for any value of $\lambda/z_0$.
In Fig. \ref{fig:regiao-y} (dipole oriented in $y$-direction), we highlight the value 
$\lambda/z_0 \approx 1.2$, above which we only have valley regime, for any
value of ${\epsilon_{2}}/{\epsilon_{1}}$.
These limits imply in the surprising behaviors illustrated in Figs. \ref{fig:regiao-epsilon101} - \ref{fig:regiao-epsilon100}. 
As completeness, in a similar way, we also investigate, in Fig. \ref{fig:regiao-z}, 
these regimes when the dipole is oriented in the $z$-direction. 
In this case, we have only transition between the peak and valley regime if ${\epsilon_{2}}/{\epsilon_{1}} > 1$, and we highlight the value $\lambda/z_0 \approx 1.74$, above
which there is no value of ${\epsilon_{2}}/{\epsilon_{1}}$ that makes the peak regime possible.

\begin{figure}[h]
\centering  
\subfigure[]{\label{fig:regime-intermediario1}\epsfig{file=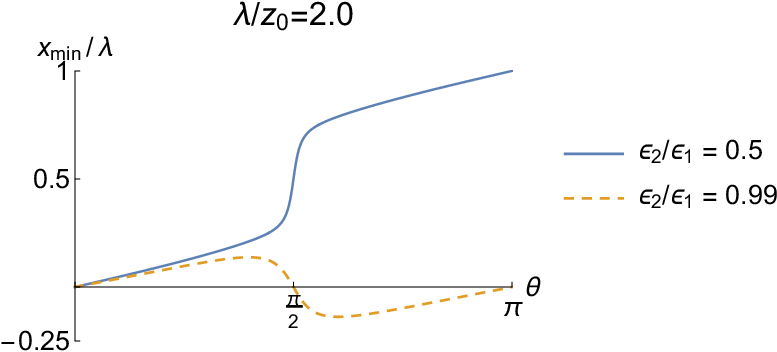, width=0.8 \linewidth}}
\hspace{4mm}
\subfigure[]{\label{fig:regime-intermediario2}\epsfig{file=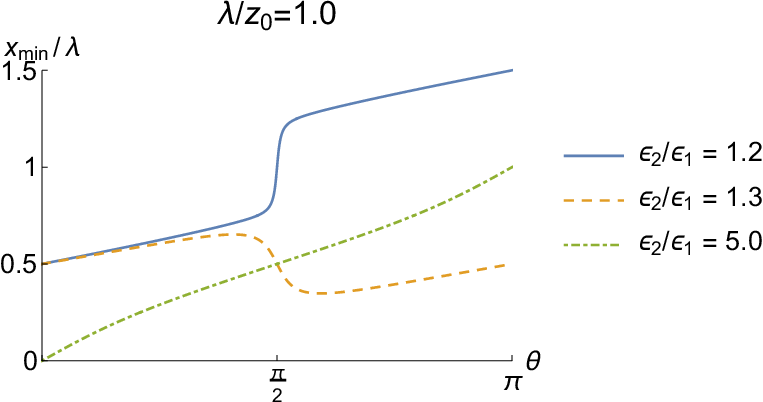, width=0.8 \linewidth}}
\caption{
Illustration of the possible behaviors for the intermediate regime by representing $ x_{\text{min}}/\lambda$ as a function of $0 \leq \theta \leq \pi$, with $\phi=0$.
In (a) we illustrate it when ${\epsilon_{2}}/{\epsilon_{1}} < 1$, considering, as examples, $\lambda/z_0 = 2$ and ${\epsilon_{2}}/{\epsilon_{1}} = 0.5$ (solid line) and ${\epsilon_{2}}/{\epsilon_{1}} = 0.99$ (dashed line).
In (b) we illustrate it when ${\epsilon_{2}}/{\epsilon_{1}} > 1$, considering, as examples, $\lambda/z_0 = 1$ and ${\epsilon_{2}}/{\epsilon_{1}} = 1.2$ (solid line), ${\epsilon_{2}}/{\epsilon_{1}} = 1.3$ (dashed line) and ${\epsilon_{2}}/{\epsilon_{1}} = 5$ (dot-dashed line).
Note that, when ${\epsilon_{2}}/{\epsilon_{1}} < 1$ [Fig. (a)], one has two possible behaviors for the intermediate regime, whereas, when ${\epsilon_{2}}/{\epsilon_{1}} > 1$ [Fig. (b)], one has three.
}
\label{fig:regime-intermediario}
\end{figure}

As discussed earlier, when the dipole has general orientations $(\phi, \theta)$, we can have the intermediate regime, and the values $x_{\text{min}}$ can be analytically computed by solving Eq. \eqref{eq:interm}.
Considering the periodicity of $U_{\text{cla}}^{(1)}$, Figs. \ref{fig:regime-intermediario1} and \ref{fig:regime-intermediario2} illustrate, in a general way, the behavior of the intermediate regime when ${\epsilon_{2}}/{\epsilon_{1}} < 1$ and ${\epsilon_{2}}/{\epsilon_{1}} > 1$, respectively.
In Fig. \ref{fig:regime-intermediario1}, one can note that we can have two possibilities (illustrated by the solid and dashed lines) for the behavior of the intermediate regime. 
The solid line shows that $n \leq x_{\text{min}}/\lambda \leq 1 + n$ (with $ n \in \mathbb{Z} $), which means that $x_{\text{min}}$ can assume any value, but, with valley regime always occurring when $\theta=\pi/2$.
The dashed line shows that $-0.25 + n < x_{\text{min}}/\lambda < 0.25 + n$, which means that $x_{\text{min}}$ is located only around the corrugation peaks.
In Fig. \ref{fig:regime-intermediario2}, one can note that we can have three possibilities (illustrated by the solid, dashed and dot-dashed lines) for the behavior of the intermediate regime. 
The solid line shows that $0.5 + n \leq x_{\text{min}}/\lambda \leq 1.5 + n$, which means that $x_{\text{min}}$ can assume any value, but, with peak regime always occurring when $\theta=\pi/2$.
The dashed line shows that $0.25 + n < x_{\text{min}}/\lambda < 0.75 + n$, which means that $x_{\text{min}}$ is located only around the valleys.
The dot-dashed line shows that $n \leq x_{\text{min}}/\lambda \leq 1 + n$, which also means that $x_{\text{min}}$ can assume any value, but now, with valley regime occurring when $\theta=\pi/2$.

It is important to remark that all these nontrivial effects of the lateral force are 
better described by our formulas as $a/z_0 \ll 1$.
Taking into account this approximation, and considering an additional one given by the limit $\lambda/z_0 \to \infty$, Eq. \eqref{eq:u1-cos} leads to
\begin{equation}
U_{\text{cla}}^{(1)} \to-a\cos\left(kx_{0}\right)\frac{3}{z_{0}}\frac{1}{\epsilon_{2}}\left(\frac{\epsilon_{1}-\epsilon_{2}}{\epsilon_{1}+\epsilon_{2}}\right)\frac{{d}_{x}^{2}+{d}_{y}^{2}+2{d}_{z}^{2}}{64\pi\epsilon_{0}z_{0}^{3}}, \label{eq:pfa1}
\end{equation}
which can also be written as
\begin{equation}
U_{\text{cla}}^{(1)} \to-h(x_{0})\frac{dU_{\text{cla}}^{(0)}(z_{0})}{dz_{0}}. \label{eq:pfa2}
\end{equation}
This result (originated by the limit $\lambda/z_0 \to \infty$) is characteristic of the PFA \cite{Clinton-PRB-1985}, and, from Eq. \eqref{eq:pfa1}, one can see that if $\epsilon_{1}>\epsilon_{2}$ the PFA just allows to see the peak regime, whereas if $\epsilon_{2}>\epsilon_{1}$ it just allows to see the valley regime.

\onecolumngrid

\twocolumngrid

\onecolumngrid

\section{Van der Waals interaction}
\label{sec-vdw}

\twocolumngrid

Let us consider a neutral polarizable particle, put at ${\bf r}_{0}=x_0\hat{{\bf x}}+y_0\hat{{\bf y}}+z_0\hat{{\bf z}}$ $(z_0>0)$, embedded in a dielectric $\epsilon_{2}$, as illustrated in Fig. \ref{fig:carga-2-meios-corrug} (with $\textbf{r}^\prime=\textbf{r}_0$).
The vdW interaction $U_\text{vdW}$ between this particle and a dielectric $\epsilon_{1}$ [Fig. \ref{fig:carga-2-meios-corrug}] can be written as $U_\text{vdW}\approx U^{(0)}_\text{vdW} + U^{(1)}_\text{vdW}$.
The first term $U^{(0)}_\text{vdW}$, which can be written as
\begin{equation}
U_{\text{vdW}}^{(0)}(z_{0})=-\frac{1}{\epsilon_{2}} \left(\frac{\epsilon_{1}-\epsilon_{2}}{\epsilon_{1}+\epsilon_{2}}\right)\frac{\langle \hat{d}_{x}^{2}\rangle+ \langle\hat{d}_{y}^{2} \rangle + 2\langle \hat{d}_{z}^{2} \rangle }{64\pi\epsilon_{0}z_{0}^{3}}, \label{U-vdW-0}
\end{equation}
is the vdW interaction energy between the particle and the dielectric $\epsilon_{1}$, for the case of a planar interface, as illustrated in Fig. \ref{fig:carga-2-meios-plano}. 
In Eq. \eqref{U-vdW-0}, one can see that, when $\epsilon_{2}<\epsilon_{1}$, one has the particle attracted towards the planar interface, whereas if $\epsilon_{2}>\epsilon_{1}$, it occurs a repulsion and a reduction of the magnitude of the force.
The second term, $ U^{(1)}_{\text{vdW}} $, which is the first-order correction to $U^{(0)}_{\text{vdW}}$ due to the corrugation in the interface between $\epsilon_{1}$ and $\epsilon_{2}$, is obtained by substituting $G^{(1)}$, given by Eq. \eqref{eq:sol-g1-coord} [with ${\cal G}^{(1)}$ given by Eqs. \eqref{eq:sol-g1}-\eqref{eq:p}], in Eq. \eqref{eq:Eberlein_Zietalquantum}, obtaining
\begin{align}
U_{\text{vdW}}^{(1)}(\mathbf{r}_{0})= & -\frac{\epsilon_{1}^2(1-\frac{\epsilon_{2}}{\epsilon_{1}})}{\epsilon_{2}\left(\epsilon_{1}+\epsilon_{2}\right)^{2}}\sum_{i,j}\frac{\langle  \hat{d}_{i}\hat{d}_{j} \rangle}{64\pi\epsilon_{0}z_{0}^{4}}\int\frac{d^{2}\mathbf{q}}{\left(2\pi\right)^{2}}\tilde{h}\left(\mathbf{q}\right)\nonumber \\
& \times e^{i\mathbf{q}\cdot{\bf r}_{0\parallel}}\left[\mathcal{I}_{ij}^{\text{cond}}\left(\mathbf{q}z_{0}\right)+\frac{\epsilon_{2}}{\epsilon_{1}}\mathcal{I}_{ij}^{\text{diel}}\left(\mathbf{q}z_{0}\right)\right], \label{eq:principal-quantico}
\end{align}
where 
$
\langle \hat{d}_{i} \hat{d}_{j}\rangle = \frac{\hbar}{\pi}\int_{0}^{\infty}d\xi\,\alpha_{ij}\left(i\xi,\epsilon_{2}\right),
$
with $\alpha_{ij}$ being the components of the polarizability tensor \cite{Buhmann-DispersionForces-I,Buhmann-DispersionForces-II}.
The dependence of the polarizability on $\epsilon_{2}$, comes from the fact that this property is modified when the particle is embedded in a dielectric medium \cite{Fiedler-JPCA-2017}.
Here, we assume that, in a first approximation, the polarization tensor is given by $\alpha_{ij}\left(i\xi,\epsilon_{2}\right)\approx f(\epsilon_{2})\alpha_{ij}^0\left(i\xi\right)$, where $\alpha_{ij}^0\left(i\xi\right)$ is the vacuum polarizability, and $f(\epsilon_{2})>0$ is a function which carries all the dependence of the polarizability on $\epsilon_{2}$.

Let us investigate the case of a sinusoidal corrugated surface with amplitude $a$ and corrugation period $\lambda$, which is described by $h(x)=a\cos(k x)$, where  $k = 2\pi/\lambda$ and $a\ll z_0$.
Then, Eq. \eqref{eq:principal-quantico} leads to
\begin{align}
U_{\text{vdW}}^{(1)}(\mathbf{r}_{0})= -\frac{\epsilon_{1}^2}{\epsilon_{2}\left(\epsilon_{1}+\epsilon_{2}\right)^{2}} & \frac{3a A(\frac{\epsilon_{2}}{\epsilon_{1}},\langle\hat{d}_i \hat{d}_j\rangle,kz_{0})}{512\pi\epsilon_{0}z_{0}^{4}} \nonumber \\
\times \cos& \left[kx_{0} -\delta\left(\frac{\epsilon_{2}}{\epsilon_{1}},\langle\hat{d}_i \hat{d}_j\rangle,kz_{0}\right)\right], \label{eq:potential-energy-CP}
\end{align}
where $A(\frac{\epsilon_{2}}{\epsilon_{1}},\langle\hat{d}_i \hat{d}_j\rangle,kz_{0})$ and $\delta(\frac{\epsilon_{2}}{\epsilon_{1}},\langle\hat{d}_i \hat{d}_j\rangle,kz_{0})$ 
are obtained by replacing 
${d}_i {d}_j \rightarrow \langle\hat{d}_i \hat{d}_j\rangle$ in Eqs. \eqref{eq:delta}-\eqref{eq:C}.
Similar to the classical case, taking the limit $\lambda/z_0 \to \infty$, one obtains
\begin{equation}
U_{\text{vdW}}^{(1)} \to-a\cos\left(kx_{0}\right)\frac{3}{z_{0}}\frac{1}{\epsilon_{2}}\left(\frac{\epsilon_{1}-\epsilon_{2}}{\epsilon_{1}+\epsilon_{2}}\right)\frac{\langle\hat{d}_{x}^{2}\rangle+\langle\hat{d}_{y}^{2}\rangle+2\langle\hat{d}_{z}^{2}\rangle}{64\pi\epsilon_{0}z_{0}^{3}}, \nonumber \\
\end{equation}
so that, one can also write
\begin{equation}
U_{\text{vdW}}^{(1)} \to -h(x_{0})\frac{dU_{\text{vdW}}^{(0)}(z_{0})}{dz_{0}}.
\end{equation}
which is characteristic of the PFA \cite{Dalvit-JPA-2008}.
One can see that this limit eliminates the presence of the phase function $\delta$, so that, the intermediate regime and the transitions between peak and valley regimes can not be predicted if this approximation is used.

\begin{figure}[h]
\centering  
\epsfig{file=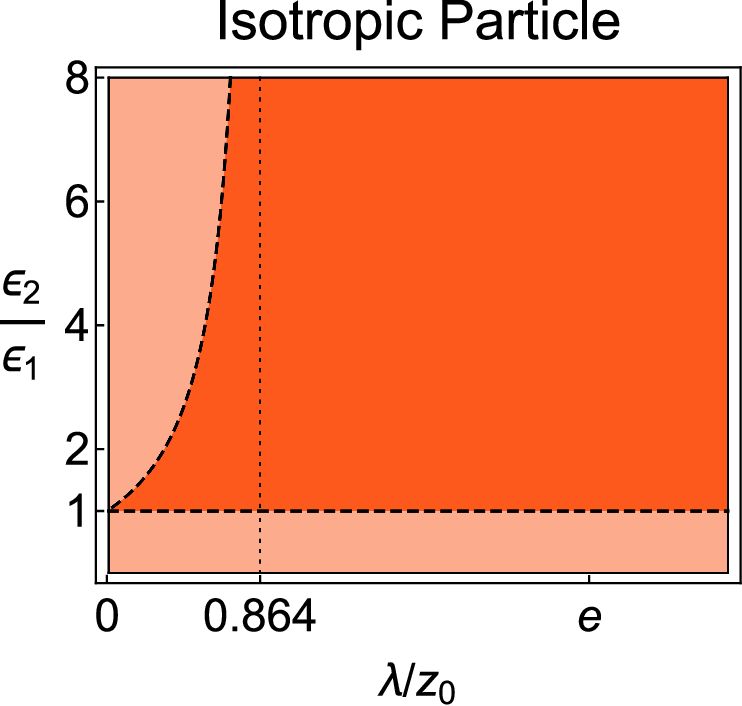, width=0.6 \linewidth}
\caption{
Illustration of the behavior of $x_{\text{min}}$ of $U^{(1)}_\text{vdW}$, represented by the configuration space ${\epsilon_{2}}/{\epsilon_{1}}$ (vertical axis) versus $\lambda/z_0$ (horizontal axis), for a general isotropic particle.
The dark region represent $C<0$, and correspond to the valley regime.
The lighter region represent $C > 0$, and correspond to the peak regime.
The border between the lighter and dark regions (dashed lines) corresponds to the situations where the lateral force vanishes $(C = 0)$.
We highlight the value $\lambda/z_0 \approx 0.864$ (dotted line), so that above it there is no value of ${\epsilon_{2}}/{\epsilon_{1}}$ that makes the peak regime possible.
}
\label{fig:regiao-isotropica}
\end{figure}

Considering isotropic particles, we have $\langle\hat{d}_x \hat{d}_z\rangle = 0$ and $\langle\hat{d}^{2}_x\rangle= \langle\hat{d}^{2}_y\rangle= \langle\hat{d}^{2}_z\rangle = \langle\hat{d}^{2}\rangle>0$, which leads to $B(\langle\hat{d}_i \hat{d}_j\rangle,kz_{0})=0$ and $C= \left(1-\frac{\epsilon_{2}}{\epsilon_{1}}\right) \langle \hat{d}^{2}\rangle \sum_{i} [\mathcal{R}_{ii}^{\text{cond}}\left(kz_{0}\right)+\frac{\epsilon_{2}}{\epsilon_{1}}\mathcal{R}_{ii}^{\text{diel}}\left(kz_{0}\right)]$.
Thus, the occurrence of peak and valley regimes depending on $\lambda/z_0$ and ${\epsilon_{2}}/{\epsilon_{1}}$ is shown in Fig. \ref{fig:regiao-isotropica}.
It is important to remark that, the sign of the function $C$ is not affected by $\langle \hat{d}^{2}\rangle$, which means the behavior illustrated in Fig. \ref{fig:regiao-isotropica} is valid to any isotropic particle.
In this figure, one can note that, for ${\epsilon_{2}}/{\epsilon_{1}} < 1$, an isotropic particle, under the action of the lateral vdW force, can only be attracted by the corrugation peaks.
Otherwise, for ${\epsilon_{2}}/{\epsilon_{1}} > 1$, one can see that a transition between peak and valley regime can occur, and we highlight the value $\lambda/z_0 \approx 0.864$, so that above it there is no value of ${\epsilon_{2}}/{\epsilon_{1}}$ that makes the peak regime possible.
The existence of two regimes when ${\epsilon_{2}}/{\epsilon_{1}} > 1$, for an isotropic particle, is very surprisingly, since that such behavior is associated, in general, with anisotropy.

\begin{figure}[h]
\centering  
\subfigure[]{\label{fig:regiao-epsilon05-quantico}\epsfig{file=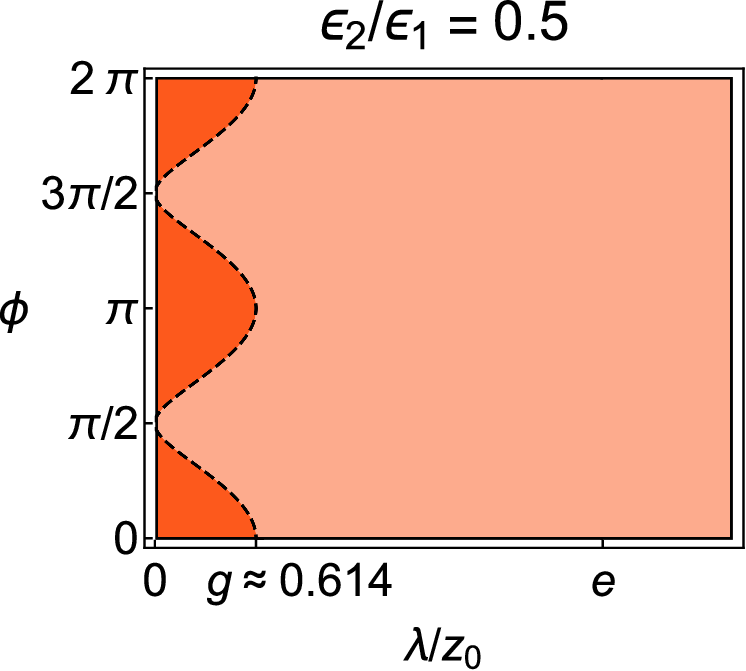, width=0.22 \linewidth}}
\hspace{4mm}
\subfigure[]{\label{fig:regiao-epsilon101-quantico}\epsfig{file=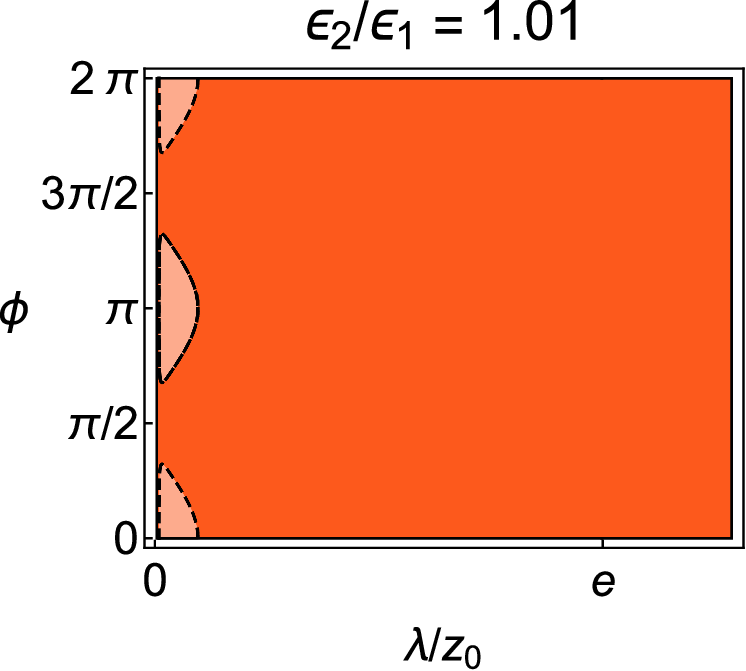, width=0.22 \linewidth}}
\hspace{4mm}
\subfigure[]{\label{fig:regiao-epsilon11-quantico}\epsfig{file=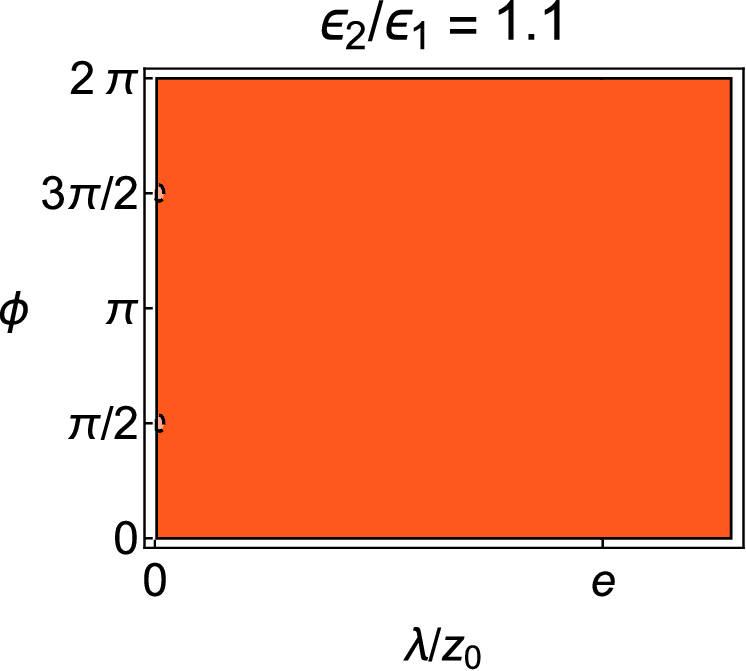, width=0.22 \linewidth}}
\hspace{4mm}
\subfigure[]{\label{fig:regiao-epsilon100-quantico}\epsfig{file=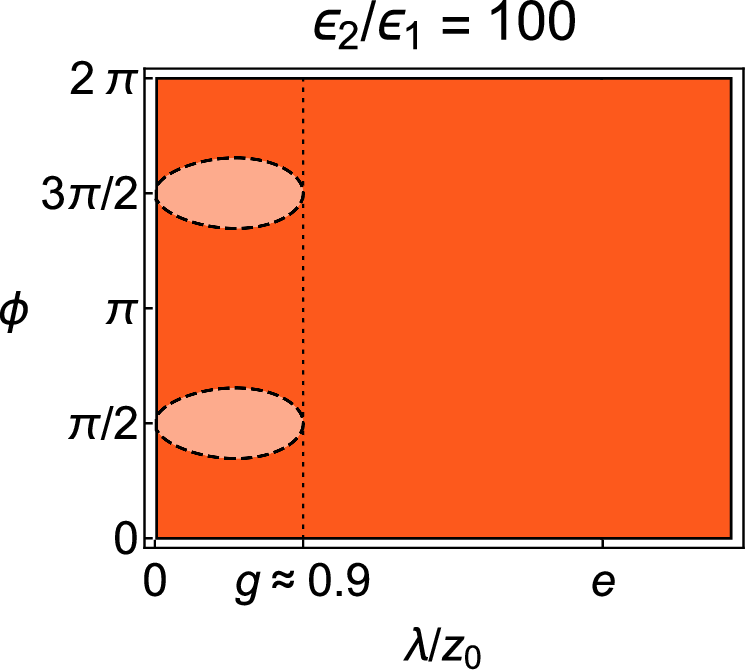, width=0.22 \linewidth}}
\caption{ 
Each figure, representing a configuration space $\phi$ (vertical axis) versus $\lambda/z_0$ (horizontal axis), for a given value of ${\epsilon_{2}}/{\epsilon_{1}}$, show the behavior of $x_{\text{min}}$ of $U^{(1)}_\text{vdW}$. 
In all these figures, 
we consider a particle characterized by $ \left\langle d_n^2 \right\rangle/\left\langle d_p^2 \right\rangle = 0.6 $ and oriented with $\theta=\pi/2$.
The dark regions represent $C(\theta=\pi/2)<0$, and correspond to the valley regime.
The lighter regions represent $C(\theta=\pi/2) > 0$, and correspond to the peak regime.
The border between the lighter and dark regions (dashed lines) corresponds to the situations where the lateral force vanishes $(C = 0)$.
The considered values for ${\epsilon_{2}}/{\epsilon_{1}}$ are: 
(a) ${\epsilon_{2}}/{\epsilon_{1}} = 0.5$ (the particular case addressed in Ref. \cite{Nogueira-PRA-2021}), 
(b) ${\epsilon_{2}}/{\epsilon_{1}} = 1.01$, 
(c) ${\epsilon_{2}}/{\epsilon_{1}} = 1.1$, 
(d) ${\epsilon_{2}}/{\epsilon_{1}} = 100$.
}
\label{fig:regiao-epsilon-quantico}
\end{figure}

Considering an anisotropic particle oriented with its principal axes coinciding with $xyz$, according to Eq. \eqref{eq:C} one has $B=0$, so that only peak [Fig. \ref{fig:regimes}(i)] or valley regime [Fig. \ref{fig:regimes}(ii)] can occur. 
This reveals behaviors of $U^{(1)}_\text{vdW}$ similar to those shown in Fig. \ref{fig:regiao-epsilon}. 
To exemplify these behaviors, let us consider a class of particles whose tensor $\left\langle d_id_j\right\rangle$ diagonalized has the form
\begin{equation} \label{eq:didj}
	\left\langle d_id_j\right\rangle=\begin{pmatrix}
		\left\langle d_n^2 \right\rangle&0&0\\
		0&\left\langle d_n^2 \right\rangle&0\\
		0&0&\left\langle d_p^2 \right\rangle
	\end{pmatrix},
\end{equation}
with $\left\langle d_p^2 \right\rangle \geq \left\langle d_n^2\right\rangle$, 
and consider the particle orientation in terms of the spherical angles $(\phi, \theta)$.
Thus, the behavior of peak and valley regime depending on $\phi$ and $\lambda/z_0$ for some values of ${\epsilon_{2}}/{\epsilon_{1}}$ is shown in Fig. \ref{fig:regiao-epsilon-quantico}, where we consider a particle characterized by $ \left\langle d_n^2 \right\rangle/\left\langle d_p^2 \right\rangle = 0.6 $ [note that, this ratio characterizes the particle in any dielectric $\epsilon_{2}$, since the approximation $\langle d_i^2 \rangle \propto f(\epsilon_{2})$ is considered].
When ${\epsilon_{2}}/{\epsilon_{1}} = 0.5$, one has Fig. \ref{fig:regiao-epsilon05-quantico}. 
Comparing this figure with the classical case shown in Fig. \ref{fig:regiao-epsilon05}, we have a very similar behavior, with the difference only in the value of $g$, which is smaller in the vdW situation.
When ${\epsilon_{2}}/{\epsilon_{1}} = 1.01$, one has Fig. \ref{fig:regiao-epsilon101-quantico}.
In the transition from Fig. \ref{fig:regiao-epsilon05-quantico} to \ref{fig:regiao-epsilon101-quantico},
we have an inversion from lighter to dark (and vice-versa) regions, and also a reduction of the values for $\phi$ related to the peak regime, as well as a general reduction of the lighter region.
Note that, in the correspondent transition in the classical case, from Figs. \ref{fig:regiao-epsilon05} to \ref{fig:regiao-epsilon101}, these reductions are less accentuated.
When ${\epsilon_{2}}/{\epsilon_{1}} = 1.1$, one has Fig. \ref{fig:regiao-epsilon11-quantico}, where we see the extinction of the possibilities of the peak regime around $\phi=0,\pi$, and an emergence of a tiny possibility of this regime around $\phi=\pi/2, 3\pi/2$.
Comparing this figure with the correspondent classical case shown in Fig. \ref{fig:regiao-epsilon11}, one can see that we, practically, do not have peak regime.
When ${\epsilon_{2}}/{\epsilon_{1}} = 100$, one has Fig. \ref{fig:regiao-epsilon100-quantico}, where we see an enhancement of the peak regime around $\phi=\pi/2, 3\pi/2$.
Comparing this figure with the correspondent classical case shown in Fig. \ref{fig:regiao-epsilon100}, one can see that we have a very similar behavior, with the difference only in the value of $g$, which is smaller in the vdW situation.
In summary, we can say that for each value of $\epsilon_{2}/\epsilon_{1}<1$, the peak and valley regimes for the classical and vdW situations are described by very similar figures, different from each other by the areas of the lighter and dark regions.
On the other hand, although for $\epsilon_{2}/\epsilon_{1}>1$ all behaviors for the peak and valley regimes for the classical case will have a similar correspondent in the vdW case, when considering specific values of $\epsilon_{2}/\epsilon_{1}>1$, these behaviors can be very different, as can be visualized comparing Figs. \ref{fig:regiao-epsilon11-quantico} and \ref{fig:regiao-epsilon11}.

\onecolumngrid

\twocolumngrid

When the particle is oriented in such a way that its principal axes do not coincide with $xyz$, one has $B \neq 0$, so that Eq. \eqref{eq:delta} leads to the presence of the phase function $\delta$ in such a way that $\delta \neq 0$ and $\delta \neq \pi$. 
This means that the intermediate regime occurs [see Fig. \ref{fig:regimes}(iii)].
Here, the behavior of the intermediate regime follows the same idea discussed in the classical case, which means that the behaviors of this regime, as a function of $\theta$, are similar to those shown in Fig. \ref{fig:regime-intermediario}.

%
\section{Final remarks}
\label{sec-final}

We studied how the peak, valley and intermediate regimes for the lateral vdW force are affected by the consideration of two non-dispersive semi-infinite dielectrics, $\epsilon_{1}$ and $\epsilon_{2}$, separated by a corrugated interface.
Our main result is given by Eq. \eqref{eq:principal-quantico}, which, applied to a sinusoidal corrugation [Eq. \eqref{eq:potential-energy-CP}], showed that when $\epsilon_{2}/\epsilon_{1} <1$
these regimes have, unless numerical factors, behaviors similar to those found for the 
situation discussed in Ref. \cite{Nogueira-PRA-2021}, where an anisotropic particle is in vacuum and interacting with a perfectly conducting surface.
For the case $\epsilon_{2}/\epsilon_{1}>1$, 
we showed that the differences from $\epsilon_{2}/\epsilon_{1}<1$ are far beyond of a mere permute between the peak and valley regimes.
For instance, when the particle is oriented in $xy$-plane, enhancing the ratio $\epsilon_{2}/\epsilon_{1}$, one has a change of the peak regimes: from the particle oriented around  $\phi=0,\pi$ to $\phi=\pi/2,3\pi/2$.
As another example, we can have situations of almost complete elimination of the peak regime.
Moreover, we showed a remarkable existence of the peak and valley regimes for an isotropic particle when ${\epsilon_{2}}/{\epsilon_{1}} > 1$ (whereas for ${\epsilon_{2}}/{\epsilon_{1}} < 1$, the existence of these regimes demands anisotropy).

In the classical context, involving a particle presenting a permanent electric dipole moment, our main result is given by Eq. \eqref{eq:principal}, with the application to a sinusoidal corrugation [Eq. \eqref{eq:u1-cos}] providing the description of the peak, valley and intermediate regimes in this context.
Eqs. \eqref{eq:u1-cos} and \eqref{eq:potential-energy-CP} provide, in classical and quantum domains, respectively, a first estimate about the non-trivial behavior of the peak, valley and intermediate regimes in the presence of dielectric media.

\begin{acknowledgments}
	L.Q. and E.C.M.N. acknowledge the support of the Coordena\c{c}\~{a}o de Aperfei\c{c}oamento de Pessoal de N\'{i}vel Superior - Brasil (CAPES), Finance Code 001.
\end{acknowledgments}
%


%

\end{document}